\documentclass[reprint,prd,12pt,superscriptaddress,onecolumn,nofootinbib]{revtex4-2}
\usepackage{hyperref}
\usepackage{ifpdf}
\usepackage{slashed}
\usepackage{subfig}
\usepackage[normalem]{ulem}
\usepackage{amsmath,amssymb,amsfonts,gensymb}
\usepackage{epsf}
\usepackage{rotating}
\usepackage{nicefrac}
\usepackage{color,xcolor,pstricks}
\usepackage{fancyhdr}
\usepackage{lineno}
\usepackage{braket}
\usepackage{multirow}
\usepackage{empheq}
\usepackage{graphics}
\usepackage{mathtools}

\definecolor{bv}{rgb}{0.54, 0.17, 0.89}
\allowdisplaybreaks

\def\bar {\overline}

\def\delacp{\Delta A_{\rm CP}}

\def\beq{\begin{equation}}
\def\eeq{\end{equation}}
\def\bea{\begin{eqnarray}}
\def\eea{\end{eqnarray}}
\def\barr{\begin{array}}
\def\earr{\end{array}}
\def\nn {\nonumber}

\usepackage{color}

\usepackage[normalem]{ulem}

\definecolor{darkgreen}{cmyk}{1,0,1,0.4}

\def\com2#1{\textcolor{red}{\textit{#1}}}

\begin{document}

\author{W. Altmannshofer}
\email{waltmann@ucsc.edu}
\affiliation{Department of Physics, University of California Santa Cruz,
and Santa Cruz Institute for Particle Physics,
1156 High St., Santa Cruz, CA 95064, US}

\author{S. Roy}	
\email{shibasis.cmi@gmail.com}
\affiliation{Chennai Mathematical Institute, Siruseri 603103, Tamil Nadu, India}

\title{A joint explanation of the \texorpdfstring{\boldmath $B\to \pi K$}{B -> pi K} puzzle and the \texorpdfstring{\boldmath $B \to K \nu \bar{\nu}$}{B -> K nu nu-bar} excess}

\begin{abstract}
In light of the recent branching fraction measurement of the $B^{+}\to K^{+} \nu\bar{\nu}$ decay by Belle II and its poor agreement with the SM expectation, we analyze the effects of an axion-like particle (ALP) in $B$ meson decays. We assume a long-lived ALP with a mass of the order of the pion mass that decays to two photons. We focus on a scenario where the ALP decay length is of the order of meters such that the ALP has a non-negligible probability to decay outside the detector volume of Belle II, mimicking the $B^{+}\to K^{+} \nu\bar{\nu}$ signal. Remarkably, such an arrangement is also relevant for the long-standing $B\to \pi K$ puzzle by noting that the measured $B^{0}\to \pi^{0}K^{0}$ and $B^{+}\to \pi^{0}K^{+}$ decays could have a $B^{0}\to a K^{0}$ and $B^{+}\to a K^{+}$ component, respectively. We also argue based on our results that the required ALP-photon effective coupling belongs to a region of parameter space that can be extensively probed in future beam dump experiments like SHiP.
\end{abstract}

\maketitle

\section{Introduction}

The presence of flavor-violating quark decays is the most significant prediction of the Cabibbo-Kobayashi-Maskawa (CKM) mechanism~\cite{Kobayashi:1973fv} within the Standard Model (SM) that was firmly established by the $B$ factories BaBar~\cite{BaBar:1998yfb, BaBar:2014omp} and Belle~\cite{Belle:2000cnh, BaBar:2014omp} as well as by LHCb~\cite{LHCb:2012myk} and Belle II~\cite{Belle-II:2018jsg} more recently. As the theory predictions and experimental observations~\cite{HFLAV:2022esi, ParticleDataGroup:2024cfk} are becoming more and more precise, flavor violating decays have come under increased scrutiny as any discrepancy can be interpreted as potential hint of physics beyond the SM (BSM). Since these decays may get affected by contributions coming from new physics above the electroweak symmetry breaking scale, they could provide indirect evidence of new heavy particles. In this context, semileptonic $B$ meson decays are particularly useful because they have clean
experimental signatures, fairly well-controlled theoretical uncertainties, and suppressed SM rates, making them sensitive probes of BSM physics. Interestingly, there exist a number of deviations from the SM predictions that involve $b\to c \,\ell^{-} \bar{\nu}_{\ell}$ and $b\to s\ell^{+}\ell^{-}$ transitions~\cite{Capdevila:2023yhq} inferred from global fits to available data.

At the same time, $B$ meson decays are also well suited to search for feebly coupled light BSM particles with masses smaller than $m_{B}$~\cite{Batell:2009jf, Freytsis:2009ct, Schmidt-Hoberg:2013hba, Dolan:2014ska, Izaguirre:2016dfi, Sala:2017ihs, Datta:2017ezo, Dror:2017ehi, Dror:2017nsg, Altmannshofer:2017bsz, Winkler:2018qyg, Dobrich:2018jyi, Aloni:2018vki, Gavela:2019wzg, Filimonova:2019tuy, MartinCamalich:2020dfe, Kachanovich:2020yhi, Chakraborty:2021wda, Dutta:2021afo, Darme:2021qzw, Bertholet:2021hjl, Bauer:2021mvw, Ferber:2022rsf, Crivellin:2022obd, Bonilla:2022qgm, Bonilla:2022vtn, He:2022ljo, Zhang:2023wlt, Ovchynnikov:2023von}. Detecting these light particles can be challenging even if they have nonzero coupling to SM particles as the light particles can be long-lived, i.e. the decay happens outside the detector volume. In terms of experimental signature, this gives rise to a signal where a $B$ meson decays for example to a kaon and missing energy. Alternatively, if the mass of the light particle is within the mass window of a known SM resonance~\cite{Ishida:2020oxl, Bhattacharya:2021ggm} then disentangling the signal from the SM background may be difficult in the case when both decay to the same final state. An illustrative example of this scenario is the two photon decay of a light particle with mass close to the pion mass. 
For the specific example of an axion-like particle (ALP) in the MeV-GeV mass range, multiple complementary search strategies have been explored employing for example collider and beam dump experiments, and searches at flavor factories~\cite{Mimasu:2014nea,Jaeckel:2015jla,Knapen:2016moh,Alves:2017avw,Dolan:2017osp,Brivio:2017ije,Bauer:2017ris,Bjorkeroth:2018dzu,Bauer:2018uxu,CidVidal:2018blh,Dobrich:2019dxc,Merlo:2019anv,Altmannshofer:2019yji,Hook:2019qoh,Gori:2020xvq,Blinov:2021say,Dreyer:2021aqd,Bandyopadhyay:2021wbb,Jerhot:2022chi,Altmannshofer:2022ckw,Acanfora:2023gzr}.

Recently, the Belle II Collaboration released an analysis~\cite{Belle-II:2023esi} of the flavor-changing neutral current (FCNC) process $B^{+}\to K^{+}\nu\bar{\nu}$ that suggests a $2.8\sigma$ excess over the predicted SM branching ratio. Since the final state neutrinos are not observed, the $B^{+}\to K^{+}\nu\bar{\nu}$ signal is similar to $B^{+}\to a K^{+}$, with $a$ escaping the detector. Moreover, it is well known that the theoretical uncertainty of the $B^{+}\to K^{+}\nu\bar{\nu}$ branching ratio prediction~\cite{Colangelo:1996ay,Buchalla:2000sk,Altmannshofer:2009ma,Buras:2014fpa,Browder:2021hbl,Bause:2021cna,Felkl:2021uxi,Becirevic:2023aov} is nominal compared to the closely related decays based on the 
$b\to s\ell^{+}\ell^{-}$ process, which may have sizeable non-local   contributions. The combination of both of these facts has resulted in a number of analyses that interpret the experimental result in the context of models with dark matter, sterile neutrinos, axion-like particles, new neutral light gauge bosons, massless dark photons, and other new FCNC-inducing light particles weakly coupled to the SM~\cite{Athron:2023hmz,Bause:2023mfe,Allwicher:2023xba,Felkl:2023ayn,Dreiner:2023cms,Abdughani:2023dlr,He:2023bnk,Berezhnoy:2023rxx,Datta:2023iln,Altmannshofer:2023hkn,McKeen:2023uzo,Fridell:2023ssf,Ho:2024cwk,Gabrielli:2024wys,Li:2024thq,Hou:2024vyw,He:2024iju,Bolton:2024egx,Marzocca:2024hua,Aghaie:2024jkj,Rosauro-Alcaraz:2024mvx,Eguren:2024oov,Buras:2024ewl,Hati:2024ppg,Wang:2024prt}.\\

In this work, we propose a simplified model with an axion-like particle that mediates the $b\to s$ flavor-changing transition and also has an effective coupling to photons. Such an ALP may decay outside the detector volume and therefore escape detection for a sufficiently long decay length in case it acquires a large boost during its production. As shown in Ref~\cite{Altmannshofer:2023hkn}, our assumptions are consistent with the hypothesis of a 2-body of decay of $B$ meson to a kaon and an invisible particle as a resolution to the $B^{+}\to K^{+}\nu\bar{\nu}$ result.

If some of the ALPs decay inside the detector, one also has an additional visible signature $B \to a K$, with $a \to \gamma \gamma$. Such a signal has been searched for~\cite{BaBar:2021ich} in the mass range $0.175< m_{a} \leq (m_{B^+}-m_{K^+})\sim 4.78\, \text{GeV} $ and away from the vicinity of $\pi^{0},\, \eta,\, \eta^{'}$-mass windows.

Remarkably, such a setup also provides a simple resolution to the puzzle involving $B\to \pi K$ decays if one considers the possibility that the ALP has a mass close the pion mass and a fraction of the reconstructed neutral pions in $B^{+}\to\pi^{0}K^{+}$ and $B^{0}\to \pi^{0}K^{0}$ are actually $a\to \gamma\gamma$ candidates. As we will see, this exploration points to a decay length of the ALP that falls within the sensitivity range of existing and future beam dump experiments. For an alternative attempt to address the $B \to \pi K$ puzzle with ALPs, see~\cite{Bhattacharya:2021shk}.
In this paper, we first strive to provide a status update of the $B\to \pi K$ puzzle in Section~\ref{Sec:I} based on the latest averages of branching fraction and $CP$-violation data involving the four $B\to \pi K$ decay modes. We also discuss how an ALP can address the puzzle. Section~\ref{Sec:III} is devoted to an overview of $B\to K$ decays with missing energy and demonstrates how a displaced ALP can jointly accommodate the $B^{+}\to K^{+}\nu\bar{\nu}$ excess events and alleviate the $B\to \pi K$ puzzle. In Section~\ref{Sec:IV}, we use the inferred ALP-photon coupling to provide an estimate of the number of ALPs that can be produced through $B$ decays and through the Primakoff process in past and future beam dump experiments, in particular CHARM and SHiP.
We conclude in Section~\ref{Sec:V}.

\section{Addressing the \texorpdfstring{\boldmath $B\to \pi K$}{B -> pi K} puzzle with axion-like particles} \label{Sec:I}

\begin{figure}[h!]
    \centering
    \includegraphics[width=0.42\linewidth]{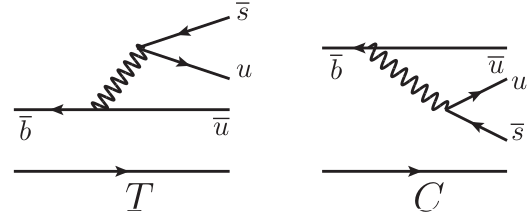} ~~~~~
    \includegraphics[width=0.52\linewidth]{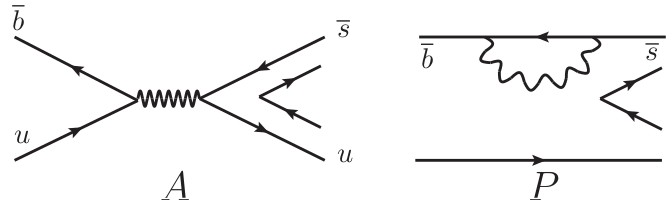}
    \caption{Topological flavor-flow amplitudes contributing to $B\to \pi K$ decays.}
    \label{fig:topo}
\end{figure}

The hadronic $B \to \pi K$ decay amplitudes can be conveniently expressed in terms of the topological flavor flow amplitudes~\cite{Gronau:1994rj, Gronau:1995hn, Fleischer:1997um, Gronau:1997an, Neubert:1998pt, Neubert:1998jq, Gronau:2006xu} $P$, $T$, $C$, $A$ illustrated in Figure~\ref{fig:topo}. These amplitudes correspond to the penguin, color-allowed tree, color-suppressed tree, and annihilation amplitudes, respectively. The penguin amplitudes are further classified into three categories,  namely the QCD penguin amplitudes $P$, color-allowed electroweak penguin (EWP) amplitudes $P_{\rm EW}$, and color-suppressed EWP amplitudes $P^C_{\rm EW}$ originating respectively from the QCD penguin operators and the electroweak penguin operator in the dim-6 hadronic effective Hamiltonian~\cite{Buchalla:1995vs,Gronau:1998fn}
 \begin{align}
\mathcal{H}_{\text{eff}}=\frac{4G_{F}}{\sqrt{2}}\Big[\lambda^{(s)}_{u}
\Big(C_{1}(O^{(u)}_{1}-O^{(c)}_{1})+C_{2}(O^{(u)}_{2}-O^{(c)}_{2})\Big)-
\lambda^{(s)}_{t} \sum_{i=1,2}C_{i}O^{(c)}_{i}-
\lambda^{(s)}_{t}
\sum_{i=3}^{10}C_{i}O_{i}^{(s)}\Big].
\label{S1Ham}
\end{align}
%
where $Q_{1-2}$ are the tree, $Q_{3-6}$ are the QCD penguin and $Q_{9},\,Q_{10}$ are the two non-negligible EWP operators in the SM. The flavor-flow amplitudes carry a strong phase and a weak phase, where only the relative phases between different amplitudes have implications on the decay observables. To disentangle the weak phase information, we factor out the CKM elements $ \lambda_{q}= V_{qb}^{*}V_{qs}$ from the flavor-flow amplitudes and note that the penguin amplitude $P$ receives contributions from all the three up-type quarks in the loop,
\begin{align}
    P=\lambda_{u}P_{u}+\lambda_{c}P_{c}+\lambda_{t}P_{t} ~,
\end{align}
that can be recast in the final form using the unitarity of the CKM matrix, 
\begin{align}
    P=\lambda_{u}(P_{u}-P_{c})+\lambda_{t}(P_{t}-P_{c}) ~.
\end{align}
The decay amplitudes~\cite{Chau:1990ay, Gronau:1994rj, Gronau:1995hn, Gronau:2005kz} for $B^0\to \pi^-K^+$, $B^+\to \pi^+ K^0$, $B^0\to \pi^0 K^0$, and $B^+\to \pi^0 K^+$, in a shorthand notation are denoted by $\mathcal{A}^{-+}$, $\mathcal{A}^{+0}$, $\mathcal{A}^{00}$, $\mathcal{A}^{0+}$ respectively and expressed as
\begin{eqnarray}
	\mathcal{A}^{-+} &=& -\lambda_{u} \left(P_{uc}+T\right) - \lambda_{t} \left(P_{tc}+\frac23 P^{C}_{EW}\right)\,,\nn\\
	\mathcal{A}^{+0} &=& \lambda_{u} \left(P_{uc}+A \right)+\lambda_{t} \left(P_{tc}-\frac{1}{3}P^{C}_{EW}\right)\,,\nn\\
	\sqrt{2}\mathcal{A}^{00} &=& \lambda_{u} \left(P_{uc}-C\right) + \lambda_{t} \left(P_{tc}-P_{EW}-\frac{1}{3}P^{C}_{EW}\right)
	\,,\nn\\ 
	\sqrt{2}\mathcal{A}^{0+} &=&-\lambda_{u} \left(P_{uc}+T+C+A\right) - \lambda_{t} \left(P_{tc}+P_{EW}+\frac{2}{3}P^{C}_{EW}
	\right)\,,\label{Bto1P}
\end{eqnarray}
where $P_{uc}$ and $P_{tc}$ are short-hand notations for $P_u - P_c$ and $P_t - P_c$.
Naively, the expected hierarchy in the magnitude among the flavor-flow amplitudes go as
\begin{align}
    \left|\lambda_{t}P_{tc}\right| \, > \, \left|\lambda_{u}T\right|\, > \, 
\left|\lambda_{u}C\right| \, > \, \left|\lambda_{u}A\right|\,,\,  \left|\lambda_{u}P_{uc}\right|\,.
\label{eq:hier}
\end{align}
Numerically, a suppression factor of the order of $\lambda \approx \sin\theta_C = 0.22$, $\theta_C$ being the Cabibbo angle, is expected at every subsequent step when these amplitudes are considered from left to right. This hierarchy is a joint effect of the magnitudes of the respective CKM elements, the extra loop suppression of the penguin amplitudes, color factors, and the small ratio of the $B$ meson decay constant to the $B$ meson mass. For example, $\lambda_u/\lambda_t \sim \lambda^2$, but $P_{tc}$ is loop-suppressed compared to $T$, conservatively estimated by an order of $\lambda$ in~\cite{Gronau:1995hm}. The ratio of $P_{tc}$ and $T$ can be as large as $0.1$ as estimated in the PQCD approach~\cite{Keum:2000ms} and differs from the numerical estimate using QCD factorization~\cite{Beneke:2000ry,Beneke:2001ev}.
It also turns out that $|C/T| \sim \lambda$~\cite{Beneke:2001ev}. 
However, even within the SM, $|C/T| \sim 0.5$ is allowed~\cite{Li:2009wba} and this ratio approaching unity~\cite{Bauer:2005kd,Huitu:2009st} is also not ruled out. 
The annihilation amplitude $A$ is suppressed by a factor of $f_{B}/m_{B}\sim 0.05 \sim \lambda^2$ when compared to $T$. The EWP amplitudes can be estimated using a relation between the tree and the EWP amplitudes with the help of the $SU(3)$-flavour symmetry of the dimension-6 weak Hamiltonian given in Eq.~\eqref{S1Ham}~\cite{Buchalla:1995vs,Gronau:1998fn} 
\beq
P_{EW} \pm P_{EW}^{C}=-\frac{3}{2}\, \frac{C_{9} \pm C_{10}} {C_{1} \pm C_{2}}\, (T \pm C)\,.
\eeq
Using the numerical values of the Wilson coefficients $C_1$, $C_2$, $C_9$, and $C_{10}$ to the leading log order at the $m_{b}$ scale~\cite{Buchalla:1995vs}, one gets
\beq
\label{T-PEW reln}
P_{EW}\sim \kappa T\,,\ \ \qquad
P_{EW}^{C}\sim \kappa C\,,
\eeq
to a good approximation, where 
\beq
\kappa=-\frac{3}{2}\, \frac{C_{9}+C_{10}}{C_{1}+C_{2}}\simeq -\frac{3}{2}\, \frac{C_{9}-C_{10}}{C_{1}-C_{2}}\simeq 
0.0135\pm 0.0012\,.
\label{eq:kappa-def}
\eeq 
%
To understand the $B\to \pi K$ puzzle it can be illuminating to consider the diagrams up to $\mathcal{O}(\lambda)$ which reduce the $B\to \pi K$ amplitudes to the following form
\begin{eqnarray}
	\mathcal{A}^{-+} &=& -\lambda_{u} T - \lambda_{t} P_{tc}\,,\nn\\
	\mathcal{A}^{+0} &=& \lambda_{t}P_{tc}\,,\nn\\
	\sqrt{2}\mathcal{A}^{00} &=&  \lambda_{t} \left(P_{tc}-P_{EW}\right)
	\,,\nn\\ 
	\sqrt{2}\mathcal{A}^{0+} &=&-\lambda_{u} T - \lambda_{t} \left(P_{tc}+P_{EW}
	\right)\,.
\end{eqnarray}

The branching ratio for a $B\to \pi K$ decay is given in terms of the corresponding decay amplitude by
\begin{align}
\label{Eq: GF out}
    \mathcal{B}(B\to \pi K) = \mathcal{B}(\pi K) = \tau_{B}\frac{p_{d}}{8\pi m_{B}^{2}}\frac{G_{F}^{2}}{2}\vert \mathcal{A} \vert^{2} ~,
\end{align}
where the Fermi coupling constant $\frac{G_{F}}{\sqrt{2}}$ is factored out of the topological amplitudes and the daughter meson momentum of the 2-body decay is expressed as $p_{d}=\frac{1}{2m_{B}}(m_{B}^{4}+m_{K}^{4}+m_{\pi}^{4}-2m_{B}^{2}m_{K}^{2}-2m_{B}^{2}m_{\pi}^{2}-2m_{\pi}^{2}m_{K}^2)^{1/2}$ in the $B$ meson rest frame. 

Direct CP asymmetries, defined as
\begin{equation}
    A_{\text{CP}}=\frac{\Gamma(B(b)\to \pi K)-\Gamma(B(\bar{b})\to \pi K)}{\Gamma(B(b)\to \pi K)+\Gamma(B(\bar{b})\to \pi K)} ~,
\end{equation}
arise in $B^0\to \pi^-K^+$ and $B^+\to \pi^0 K^+$ because of the $T$-$P_{tc}$ interference leading to a nonzero relative strong phase, as well as a weak phase difference between the two topological amplitudes. 
In contrast, $P_{EW}$ and $T$ carry the same strong phase as they are related to good approximation by a real number as shown in Eq.~\eqref{eq:kappa-def}. Therefore, $P_{EW}$-$T$ interference for $B^+\to \pi^0K^+$ does not contribute to $A_{\rm CP}$. Thus, one expects a simplified relation~\cite{Gronau:1998ep}
\beq
	A_{\rm CP}(B^{0}\to \pi^{-}K^{+})=A_{\rm CP}(B^{+}\to \pi^{0}K^{+}) ~.
\eeq
Experimentally, a deviation, numerically expressed by the quantity $\delacp$
\begin{align}\label{eq:delACP}
\Delta A_{\rm CP}=A_{\rm CP}(B^{+}\to \pi^{0}K^{+})-A_{\rm CP}(B^{0}\to \pi^{-}K^{+}),
\end{align}
is found to be nonzero~\cite{Duh:2012ie, Aaij:2020wnj, Belle-II:2023ksq}. In addition, also the relative sign of $A_{\text{CP}}$ of the two decay modes is in contradiction with the SM expectation. It is instructive to express the four $B\to \pi K$ rate asymmetries, $\Delta(\pi K)=A_{\text{CP}}(\pi K)\Gamma(\pi K)$, in terms of the topological amplitudes
\begin{align}
\label{Eq:deca-rate-asym1}
	\Delta (\pi^- K^+) =& -4\, \text{Im}(\lambda_{u}^{*}\lambda_{t})\, 
	\text{Im}[(T+P_{uc})^{*}(P_{tc}+\textstyle{\frac23}P^{C}_{EW})]\, , \\ \label{Eq:deca-rate-asym2}
	2\Delta (\pi^0 K^+) =& -4\, \text{Im}(\lambda_{u}^{*}\lambda_{t})\, \text{Im}[(T+C+A+P_{uc})^{*}(P_{tc}+P_{EW}
	+\textstyle{\frac{2}{3}}P^{C}_{EW})]\, ,\\ \label{Eq:deca-rate-asym3}
	2\Delta (\pi^0 K^0) =& -4\, \text{Im}(\lambda_{u}^{*}\lambda_{t})\, \text{Im}[(P_{uc}-C)^{*}(P_{tc}-P_{EW}
	-\textstyle{\frac13}P^{C}_{EW})]\, , \\ \label{Eq:deca-rate-asym4}
	\Delta (\pi^+ K^0) =& -4\, \text{Im}(\lambda_{u}^{*}\lambda_{t})\, \text{Im}[(A+P_{uc})^{*}(P_{tc}
	-\textstyle{\frac13}P^{C}_{EW})]\, . 
\end{align}
and retain only the color-suppressed amplitude $C$ as the leading subdominant flavor-flow amplitude. With the additional assumptions that the annihilation amplitude $A$ is suppressed relative to the color-allowed tree amplitude $T$ and the relative strong phase difference between the $T$ and $C$ amplitudes is small, from Eqs.~\eqref{Eq:deca-rate-asym1}-\eqref{Eq:deca-rate-asym4} one infers a more theoretically robust CP sum rule relation~\cite{Gronau:2005kz} connecting all the four $B \to \pi K$ CP asymmetries in Eq.~\eqref{isospin-sumrule} 
\begin{eqnarray}
	\label{isospin-sumrule}
	\nn && \Delta_{4} = A_{\rm CP}(\pi^{-} K^{+})+A_{\rm CP}(\pi^{+} K^{0}) \frac{\mathcal{B}(\pi^{+} K^{0})\tau_{0}}
	{\mathcal{B}(\pi^{-} K^{+})\tau_{+}} 
\\
&& \qquad - A_{\rm CP}(\pi^{0} K^{+})\frac{2\mathcal{B}(\pi^{0} K^{+})\tau_{0}}{\mathcal{B}(\pi^{-} K^{+})\tau_{+}} - 
A_{\rm CP}(\pi^{0} K^{0})\frac{2\mathcal{B}(\pi^{0} K^{0})}{\mathcal{B}(\pi^{-} K^{+})}\,,
\end{eqnarray}
that vanishes and the result holds up to a few percent where $\mathcal{B}(\pi K)$ are the branching ratios introduced earlier and $\tau_{+}$ and $\tau_{0}$ are
the lifetimes of the $B^{+}$ and $B^{0}$ mesons, respectively. It is interesting to note that an early measurement of $\Delta_{4}=-0.270 \pm 0.132 \pm 0.060$ at Belle~\cite{Duh:2012ie} deviated from zero (albeit with large errors), while a recent measurement $\Delta_{4}=-0.03 \pm 0.13 \pm 0.04$ at Belle~II~\cite{Belle-II:2023ksq} is perfectly compatible with zero.

Some comments are in order. Firstly, a large annihilation contribution to the $B\to \pi K$ decays is an unexpected solution to the $B\to \pi K$ puzzle, as the branching fractions do not favor a sizable annihilation amplitude~\cite{Beneke:2001ev, Bobeth:2014rra, Huber:2021cgk} which is also supported by model-independent $B\to P P$ fits to available data~\cite{Baek:2004rp, Datta:2004re}. Secondly, long-distance re-scattering effects~\cite{Neubert:1997wb, Atwood:1997iw, Buras:1997cv, Buras:2000gc, Falk:1998wc, Cheng:2004ru} can not only contribute to sizable strong phases but also modify the naive hierarchy in relative size for the topological amplitudes~\cite{Buras:1998ra, Beneke:2000ry, Bauer:2004ck, Bauer:2004tj, Bauer:2005kd, Beneke:2001ev, Huitu:2009st} but is not expected to entirely resolve~\cite{Kamal:1999rn, Cheng:2005bg} the $B\to \pi K$ puzzle. (See also~\cite{Ciuchini:2001gv, Ciuchini:2008eh} for a discussion on charming penguin resolution to the $B\to \pi K$ puzzle and \cite{Beaudry:2017gtw, Fleischer:2017vrb, Fleischer:2018bld, Kundu:2021emt} for an up-to-date analysis). Therefore, following~\cite{Buras:1998ra, Beneke:2000ry, Keum:2000wi, Beneke:2003zv, Beneke:2001ev, Chang:2008tf, Cheng:2009eg, Li:2009wba, Li:2014haa, Liu:2015upa, Cheng:2009eg, Li:2005kt} we ignore both the $P_{uc}$ and $A$ contributions while allowing a non-negligible color-suppressed tree contribution to the $B\to \pi K$ decays and treat the strong phases as free parameters in the SM as expressed in Eq.~\eqref{Bto1P}
\begin{eqnarray} \label{Bto1P_1}
	\mathcal{A}^{-+} &=& -\lambda_{u} T - \lambda_{t} \left(P_{tc}+\frac23 P^{C}_{EW}\right)\,, \\
	\mathcal{A}^{+0} &=&\lambda_{t} \left(P_{tc}-\frac{1}{3}P^{C}_{EW}\right)\,, \\
	\sqrt{2}\mathcal{A}^{00} &=& -\lambda_{u}C + \lambda_{t} \left(P_{tc}-P_{EW}-\frac{1}{3}P^{C}_{EW}\right)
	\,, \\ 
	\sqrt{2}\mathcal{A}^{0+} &=&-\lambda_{u} \left(T+C\right) - \lambda_{t} \left(P_{tc}+P_{EW}+\frac{2}{3}P^{C}_{EW}
	\right)\,.\label{Bto1P_4}
\end{eqnarray}
We make use of the PDG averaged data~\cite{ParticleDataGroup:2024cfk} consisting of the four branching ratios for the $B\to \pi K$ modes, the four direct CP asymmetries $A_{\rm CP}$, and the mixing-induced CP asymmetry $S_{\rm CP}$ measured for the $B\to \pi^{0}K^{0}$ decay from BaBar, Belle, Belle~II and LHCb~\cite{Belle-II:2023ksq, Aaij:2020wnj} and perform a fit to the magnitude and strong phases contributing to Eqs.~\eqref{Bto1P_1}-\eqref{Bto1P_4}. The input data is summarized in Table~\ref{Tab:tab1}.

\begin{table}[tb]
\centering
\setlength{\tabcolsep}{12pt}
\renewcommand*{\arraystretch}{1.3}	
\begin{tabular}{c|ccc}	
\hline \hline
			Modes& Avg. BR $[10^{-6}]$	& Avg. $A_{\rm CP}$	 &$S_{\rm CP}$\\ 
			\hline
			$B^{0}\to \pi^{-}K^{+}$		&		$20.00\pm 0.04$			  &	        $-0.0831\pm 0.0031$	&	\\
			\hline
			$B^{+}\to \pi^{0}K^{+}$	    &		$13.2\pm 0.4$             &			$0.027\pm 0.012 $	&	\\
			\hline
			$B^{+}\to \pi^{+}K^{0}$	    &		$23.9 \pm 0.6$	          &	        $-0.003 \pm 0.015 $	&	\\
			\hline
			$B^{0}\to \pi^{0}K^{0}$		&	    $10.1\pm 0.4$	          &	        $0.00\pm 0.08 $	&	$0.64 \pm 0.13$	\\
\hline\hline
\end{tabular}
\caption{Experimental inputs used in this work taken from the PDG~\cite{ParticleDataGroup:2024cfk}.}
\label{Tab:tab1}
\end{table}

We have five free fit parameters: the three magnitudes $P_{tc}$, $\vert T \vert,\,\vert C \vert$ and two relative phases $\delta_{T}, \, \delta_{C}$. The parameter $\kappa$ is treated as a prior around the central value. We have defined the relative phases with respect to the $P_{tc}$ diagram whose absolute phase is set to zero in this convention~\cite{Kim:2007kx}.

Apart from these free parameters, the magnitude of the CKM elements $\vert V_{ub} \vert$, $\vert V_{us} \vert$, $\vert V_{tb} \vert$, 
$\vert V_{ts} \vert$, and the CKM angles $\beta$ and $\gamma$, enter the analysis as uncertain theoretical inputs. We use the HFLAV averages~\cite{HFLAV:2022esi} that are collected in Table~\ref{Tab:tab1a}. These values are incorporated as SM priors while the uncertainties of the meson masses and the $B$ meson lifetimes are neglected in the current analysis. It is imperative to note that the input parameters $\beta$ and $\gamma$ are correlated through the CKM unitarity relation~\cite{Buras:2002yj}
\begin{align}
    \cot{\beta}=\frac{1-\bigg\vert \frac{V_{ud}V_{ub}^{*}}{V_{cd}V_{cb}^{*}} \bigg\vert \cos{\gamma}}{\bigg\vert\frac{ V_{ud}V_{ub}^{*}}{V_{cd}V_{cb}^{*}} \bigg\vert \sin{\gamma}} ~,
\end{align}
and therefore should not be used as independent parameters. However, through explicit checks, we confirmed that the fit results are insensitive to this choice.

\begin{table}[tb]
\centering
\setlength{\tabcolsep}{12pt}
\renewcommand*{\arraystretch}{1.3}	
		\begin{tabular}{c|c}	
  \hline \hline
			Parameter& Value\\ 
			\hline
			$\vert V_{us}\vert$		&		$0.2245\pm 0.0008$		\\
            $\vert V_{ub}\vert$     &       $(3.82 \pm 0.24)\times10^{-3}$\\
            $\vert V_{ts}\vert$     &       $(41.5 \pm 0.9)\times10^{-3}$\\
            $\gamma$     &       $(65.9 \pm 3.3 \pm 3.5)^{\degree}$\\
             $\beta$     &       $(22.14 \pm 0.69 \pm 0.67)^{\degree}$\\
   \hline\hline
\end{tabular}
\caption{Relevant theoretical CKM input parameters from HFLAV averages~\cite{HFLAV:2022esi}.}
\label{Tab:tab1a}
\end{table}
\begin{table}[tb]
\centering
\setlength{\tabcolsep}{12pt}
\renewcommand*{\arraystretch}{1.3}	
\begin{tabular}{c|cccc}
\hline\hline
Scenario & $\chi^{2}/d.o.f$ &$p$-value & Fit parameter & Fit value\\
\hline
 &  & &$P_{tc}$  & $-0.149\pm 0.001$\\
\textbf{I}&1.4 &0.23 & $\kappa$ & $0.0134\pm 0.001$ \\
(SM fit)& &  &$\vert T \vert $ & $0.84\pm 0.20$\\
 &  & &$\vert C \vert $ & $0.41\pm 0.09$\\
 &  & &$\delta_{T} $ & $3.48\pm 0.09$\\
 &  & &$\delta_{C} $ & $1.14\pm 0.45$\\
\hline
& &  &$P_{tc}$ & $-0.148\pm 0.001$\\
\textbf{II}&0.75 &0.52 & $\kappa$ & $0.0130\pm 0.001$ \\
(SM + ALP) &   & & $ \vert T \vert $ & $0.98\pm 0.20$\\
&    & &$\vert C \vert $ & $0.32\pm 0.03$\\
&  & &$\delta_{T} $ & $3.42\pm 0.06$\\
&   & &$\delta_{C} $ & $1.46\pm 0.42$\\
&   & &   $\mathcal B(B^{+}\to a K^{+})\vert_{\text{mistag}}$  & $(6.50^{+3.92}_{-3.00}) \times 10^{-7}$ \\
\hline\hline
\end{tabular}
\caption{Best fit values for the topological flavor-flow amplitudes and strong phases. Scenario~\textbf{I}: SM-fit, Scenario~\textbf{II}: Fit in the presence of a neutral pseudoscalar mimicking a $\pi^{0}$. The topological amplitudes are defined by factoring out $G_{F}/\sqrt{2}$ as indicated in Eq.~\eqref{Eq: GF out} and the amplitudes are therefore given in units of $\text{GeV}^{3}$. The branching ratio of mistagged $B^{+}\to \pi^{0}K^{+}$ that alleviates the $B\to \pi K$ puzzle is quoted as $\mathcal B(B^{+}\to a K^{+})\vert_{\text{mistag}}$.
}
\label{tab:tab2}	
\end{table}

The results of the Standard Model fit are given in Scenario I of Table~\ref{tab:tab2}. While the $\chi^2$ per degree of freedom and the $p$-value are not unacceptable, the SM fit prefers a somewhat large value of $C/T \simeq 0.5$. Interestingly, if the CKM angle $\gamma$ is set to the recent measured value of $78.6^{\degree +7.2}_{-7.3}$ at Belle~II~\cite{Belle-II:2024eob}, we find that the $C/T$ ratio shifting towards even larger value of $0.68$ with increased $\Delta \chi^{2}/{d.o.f}=1.47$ and a reduced $p$-value of 0.20.

In our attempt to obtain a $C/T$ ratio consistent with the SM expectation, we postulate that an axion-like particle with a mass close to that of a pion is produced in the $B \to K$ transition and that it subsequently decays to two photons mimicking a neutral pion decay. This implies that the measured $B^{+}\to \pi^{0}K^{+}$ and  $B^{0}\to \pi^{0}K^{0}$ have $B^{+}\to a K^{+}$ and $B^{0}\to aK^{0}$ components. The experimentally observed branching fraction of $B^{+}\to \pi^{0}K^{+}$ and $B^{0}\to \pi^{0}K^{0}$ is therefore expressed as,
\begin{align}
    \mathcal B(B^{+}\to \pi^{0}K^{+})\vert_{\text{exp}}=\mathcal B(B^{+}\to \pi^{0}K^{+})+\mathcal B(B^{+}\to a K^{+}),\nonumber\\
    \mathcal B(B^{0}\to \pi^{0}K^{0})\vert_{\text{exp}}=\mathcal B(B^{0}\to \pi^{0}K^{0})+\mathcal B(B^{0}\to a K^{0}) ~,
\end{align}
and a similar relation holds for the $CP$-conjugate modes. We do not consider the ALP-neutral pion mixing in this work as the dominant mechanism for its decay to two photons in contrast to Ref.~\cite{Altmannshofer:2019yji, Bhattacharya:2021ggm} and therefore are not susceptible to the bounds on the ALP-pion mixing angle from kaon decays (see, e.g.~\cite{Alves:2017avw}). In our setup, the ALPs are long lived and can in principle be distinguished from the $B\to \pi K$ decays. We therefore do not need to consider the interference between the $B\to \pi K$ and $B\to a K$ decay amplitudes. We also ignore potential new sources of $CP$-violation in the $B\to aK$ decays.

Since we are interested in the $b\to s a$ transition, the relevant flavor-changing interaction terms are parameterized as,
\begin{align}
\label{Eq:ALP lag}
\mathcal{L}_{\text{FCNC}}\supset \overline{s}(h^{S}_{sb}+h^{P}_{sb}\gamma_{5})b \, a +\text{h.c.}
\end{align}
where $a$ is the ALP field. We note that the existence of such a term in the Lagrangian necessitates introduction of additional new states and interactions in order to couple SM quarks to a pseudoscalar singlet in a gauge-invariant way such that the couplings between the pseudoscalar and quarks actually arise from higher-dimensional operators. At the moment, we interpret Eq.~\eqref{Eq:ALP lag} as low-energy limit of a more complete theory and do not attempt to provide a complete description of that full theory. For our purpose, we use Eq.~\eqref{Eq:ALP lag} to obtain the decay rate for $B\to a K$ (see e.g.~\cite{ Batell:2009jf, Dolan:2014ska, Dolan:2017osp, Dobrich:2018jyi, Bauer:2021mvw})
\begin{align}
\label{Eq:BAK dec rate}
  \mathcal B(B\to a \, K)=\tau_{B}\frac{p_{K}}{8\pi m_{B}^{2}} \frac{(m_{B}^{2}-m_{K}^{2})^{2}}{(m_{b}-m_{s})^{2}}   \vert f^{B \to K}_{0}(m_a^2) \vert^{2} \vert h^{S}_{sb} \vert^{2}.
\end{align}
where $p_{K}=\frac{1}{2m_{B}}\lambda^{1/2}(m_{B}^{2},m_{K}^{2},m_{a}^{2})$
is the momentum of the kaon with $\lambda(x,y,z) = x^{2}+y^{2}+z^{2}-2xy-2yz-2zx$, and $f^{B \to K}_{0}$ is the form factor relevant for the $B\to K$ transition. For our purposes, it is sufficient to approximate this form factor as \cite{Ball:2004ye} (see~\cite{Gubernari:2023puw} for a recent state-of-the-art evaluation)
\begin{align}
   f_{0}^{B\to K}(q^{2})=\frac{0.33}{1-\frac{q^{2}}{38 \text{GeV}^{2}}} ~.
\end{align}
The results of the fit where the ALP mediated $b\to sa$ transition is modeled using a single parameter $h^{S}_{sb}$ in addition to the SM fit parameters are given in Scenario II of Table~\ref{tab:tab2}. Fit results in Scenario II indicate that the $\vert C \vert/\vert T \vert$ ratio prefers a lower value $< \, 0.3$ with a significant reduction in $\Delta \chi^{2}/{d.o.f}=0.75$ while the $p$-value increases to 0.52 compared to the SM fit. The best fit value of $h^{S}_{sb}$ yields a mistagged branching ratio of
\begin{align} \label{eq:BR_mistag}
    \mathcal B(B^{+}\to a K^{+})\vert_{\text{mistag}}=6.50^{+3.92}_{-3.00} \times 10^{-7} ~.
\end{align}
It is worth to keep in mind that the inferred branching fractions of $B^{+}\to a K^{+}$ and $B^{0}\to a K^{0}$ from the fits to $B \to \pi K$ data need not coincide with the true $B^{+}\to a K^{+}$ and $B^{0}\to a K^{0}$ branching fractions in the case the $a\to \gamma \gamma$ decay is displaced. An ALP decaying to $\gamma \gamma$ via an effective ALP-photon interaction $g_{a\gamma\gamma}$ has a decay rate given by
\begin{align} \label{eq:G_gg}
    \Gamma(a \to \gamma \gamma)=\frac{g_{a \gamma \gamma}^{2}}{64\pi}m_{a}^{3} ~.
\end{align}
The effective interaction $g_{a \gamma \gamma}$ therefore determines the lifetime of the ALP given by $\tau_{0}=\Gamma^{-1}$. For our assumed mass of the ALP of the order of the pion mass $m_\pi \simeq 130 \, \text{MeV}$ and $g_{a\gamma\gamma} \ll \text{1 TeV}^{-1}$, the decay length can be macroscopic, and an ALP that is produced in the lab frame with a large boost may decay outside a detector such as Belle~II and give a missing energy signature. Therefore, a potential $B\to a K$ event in Belle~II can either provide a signature that resembles $B \to \pi^0 K$ (if the ALP decays shortly after the $B$ decay) or $B \to K \nu \bar{\nu}$ (if the ALP decays outside the detector). Both signatures can occur with an $\mathcal O(1)$ probability if the ALP decay length approximately matches the size of the Belle~II detector. 

Interestingly, Belle II has recently measured~\cite{Belle-II:2023esi} a branching fraction for the $B^{+}\to K^{+}\nu \bar{\nu}$ decay which is somewhat higher than the SM expectation. 
As the $B^{+}\to K^{+}\nu \bar{\nu}$ decay and the $B^{+}\to a K^{+}$ decay with the $a\to \gamma \gamma$ decay outside the detector volume have the same visible signature, it has been suggested that the $B^{+}\to K^{+}\nu \bar{\nu}$ excess events may be explained in terms of new long-lived particles.

\section{Long-Lived ALPs and \texorpdfstring{\boldmath $B \to K \nu\bar\nu$}{B -> K nu nu-bar}} \label{Sec:III}

In this section, we investigate the role of $B \to a K$ decays as a potential source of contamination to $B^+ \to K^{+} \nu \bar{\nu}$ signal when the ALP decays invisibly or escapes the detector before decaying. In this context, the ALP must be light with mass in the range $m_{a}\leq m_{B}-m_{K}$. Additionally, the on-shell production of such a particle in the above-mentioned mass range would result in a resonant feature in the missing invariant mass spectrum. This fact has important consequences for the recent branching fraction measurement of $B^{+} \to K^{+}\nu \bar{\nu}$ at Belle II~\cite{Belle-II:2023esi}
\begin{align}
\mathcal B(B^+\to K^+ \nu \bar{\nu})=(2.3 \pm 0.5 \,^{+0.5}_{-0.4} ) \times 10^{-5} ~,
\end{align}
where an excess of $2.7 \sigma$ is observed from the SM expectation~\cite{Parrott:2022smq}
\begin{align}
\mathcal B(B^+\to K^+ \nu \bar{\nu})_\text{SM} = (5.58 \pm 0.37)\times 10^{-6} ~.
\end{align}
In order to interpret a $B^+ \to K^{+} \slashed{E}$ signal as an incoherent sum of the branching fractions of $B^+ \to K^{+}\nu \bar{\nu}$ and $B \to a K^{+}$  we need to recast the $B^+ \to K^{+} \slashed{E}$ experimental information in the kinematic regions
corresponding to an ALP with mass $m_{a}\simeq m_{\pi^{0}}$. Therefore, we should consider $B^{+} \to K^{+} \nu \bar{\nu}$ events exclusively from the $q^{2} \simeq m_{\pi^{0}}^{2}$ bin.
In a recent analysis~\cite{Altmannshofer:2023hkn}, the measured $B^{+}\to K^{+}\nu\bar{\nu}$ decay is interpreted to be a signal for $B\to K \, X$ as a function of the mass of the particle $X$. In the region where $m_{X}\leq 500\,\text{MeV}$, the branching ratio is found to be~\cite{Altmannshofer:2023hkn}
\begin{align} \label{eq:BKX}
    \mathcal B(B^{+}\to K^{+}X)\vert_{m_{X}\leq 500\, \text{MeV}}=(0.4\pm 0.2)\times 10^{-5} ~.
\end{align}
We use this estimate as the branching fraction for $B^{+}\to a K^{+}$. However, in order to interpret this branching fraction to coincide with the theoretical estimate for a hypothetical ALP produced in a $B^{+}\to a K^{+}$ decay which then escapes the detector undetected, we need to calculate the probability of such an ALP to remain stable within the detector volume. The probability of the ALP to not decay within the detector is given by
%
\begin{align}
    \text{Prob}(l\geq l_{\text{max}})=\exp{\Big[-\frac{l_{\text{max}}}{l_{a}}\Big]} ~,
\end{align}
where $l_{\text{max}}$ is the maximum distance of the displaced ALP decay vertex from the point of production that can be unambiguously reconstructed in the detector. For our calculation, we assume this length ($l_{\text{max}}$) to be of the order of 2 meters which is the longitudinal distance from the primary point of interaction to the forward endcap region of the Belle II electromagnetic calorimieter~\cite{Belle-II:2018jsg}. The ALP decay length $l_{a}$ is determined in the lab frame from its momentum $p_{a}^{\text{lab}}$ and proper lifetime of $\tau_{0}$ 
\begin{align}
    l_{a}=\beta_{a}^{\text{lab}}\gamma_{a}^{\text{lab}}c\tau_{0}=\frac{\vert p_{a}^{\text{lab}} \vert}{m_{a}}c\tau_{0} ~.
\end{align}
%
The initial $B$ meson produced in Belle II is longitudinally boosted in the lab frame which results in the majority of ALPs produced in
$B$ decays to likely propagate in the forward direction in the lab frame. The fraction of ALPs produced in $B$ decays which escape the Belle II detector longitudinal length is given by the quantity $(1-f_{L})$ where $f_{L}$ denotes the probability that an ALP decays within the detector volume yielding a sufficiently high quality vertex. The expression for $f_{L}$ can be written as 
\begin{align}
 f_{L}(m_{B},m_{K},m_{a}\simeq m_{\pi^{0}},l_{\text{max}})=\int_{0}^{\pi/2} \sin{\theta_{a}} d\theta_{a} \Big(1-\exp{\big(-\frac{m_{a}l_{max}}{c\tau_{0}\vert p_{aL}^{\text{lab}} \vert}\big)}\Big)  ~,
   \label{eq: survival probability int}
\end{align}
where $p_{aL}^{\text{lab}}$ is the longitudinal component of the ALP momentum. 

In the limit where the ALP decays just outside the decay volume we postulate,
\begin{align}
    \mathcal B(B^{+}\to a K^{+})\vert_{\text{theo}} \times (1- f_{L}) \simeq \mathcal B (B^{+}\to a K^{+})\vert_{\text{exp}}\simeq \mathcal B(B^{+}\to K^{+} \nu \bar{\nu})\vert_{q^{2}=m_{\pi^{0}}^{2}}^\text{exp} ~,
\end{align}
with $\mathcal B(B^{+}\to K^{+} \nu \bar{\nu})\vert_{q^{2}=m_{\pi^{0}}^{2}}^\text{exp}$ approximately given by the value in Eq.~\eqref{eq:BKX}.

\begin{figure}
    \centering
    \includegraphics[width=0.75\linewidth]{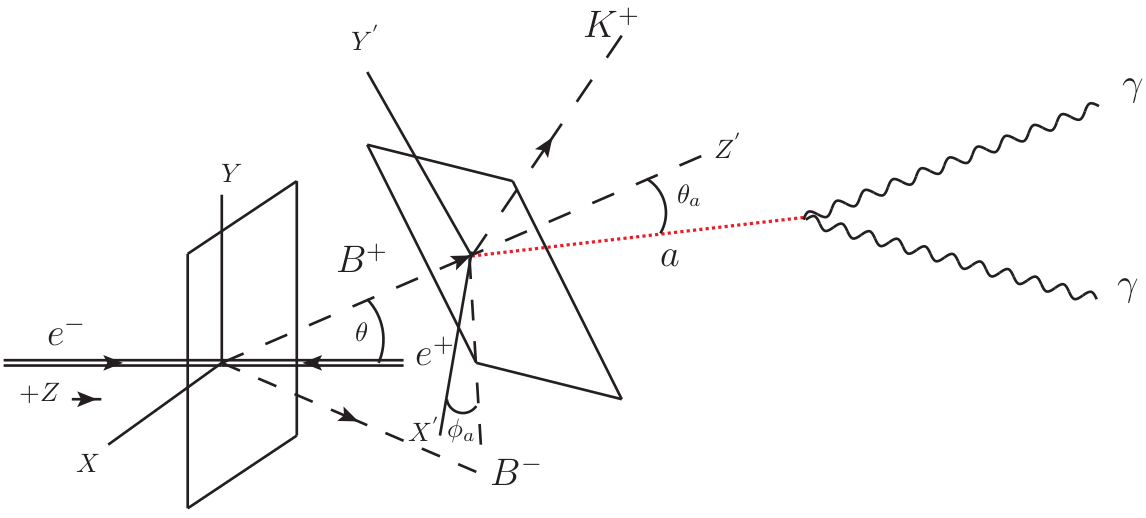}
    \caption{Production of an ALP in $B$ decays.}
    \label{fig:ALP B}
\end{figure}

On the other hand, a resolution to the $B\to \pi K$ puzzle requires some of the $B^{+,0}\to a\, K^{+,0}$ decays to be misidentified as $B^{+}\to \pi^{0}K^{+}$ and $B^{0}\to \pi^{0}K^{0}$ decays respectively, which necessitates the ALP to decay to two photons that can be reconstructed in the detector
\begin{align}
    \mathcal B(B^{+}\to a K^{+})\vert_{\text{theo}} \times f_L \simeq \mathcal B(B^{+}\to a K^{+})\vert_{\text{mistag}} ~,
\end{align}
yielding a value for $\mathcal B(B^{+}\to a K^{+})\vert_{\text{mistag}}$ given in Eq.~\eqref{eq:BR_mistag}. We have once again assumed as the limiting case that the two photons can be resolved by the detector originating from the ALP decay vertex displaced all the way to $l_{max}$ from the point of production in order to be identified as a neutral pion.  
The initial $B$ mesons are boosted in the $+z$-axis along the $e^{-}$ direction with $\gamma_{B}\beta_{B}\approx 0.28$~\cite{Filimonova:2019tuy}. The subsequent ALP decay, coming from $B \to a K$ decay in the most general case makes a polar angle $\theta_{a}$ and an azimuthal angle $\phi_{a}$ with the direction of the $B$ meson . The $B$ meson itself flies at an unknown angle $\theta$ with respect to the asymmetric beam.  Focusing on the longitudinal distance ($l_{aL}$) traveled by the ALP before decaying to two photons as shown in Fig.~\ref{fig:ALP B} we find,
\begin{align}   
   l_{aL}=\frac{c\tau_{0}\,p_{aL}^{\text{lab}}}{m_{a}};\quad p_{aL}^{\text{lab}}\,=\,\big(\gamma_{B}(p_{a}\cos{\theta_{a}}+\beta_{B} E_{a})\cos{\theta}+p_{a}\sin{\theta_{a}}\cos{\phi_{a}}\sin{\theta}\big) ~.
\end{align}

The analytical results required to evaluate Eq.~\eqref{eq: survival probability int} are provided in Appendix~\ref{App:1}. We assume the ALP dominantly decays to two photons and to no other SM particles. In principle, from phase space considerations, the ALP can decay to $e^{+}e^{-}$ or $\nu\bar{\nu}$ as well. In order to connect the ALP-photon coupling to ALP-electron coupling once again requires a concrete description of the full theory which has been described elsewhere (see e.g.~\cite{Bauer:2020jbp,Bauer:2021mvw}). In our chosen setup, the expected event rate for the ALP produced in $B^{+} \to a K^{+}$ decay and its subsequent decay to two photons is given by 
\begin{align}
    N_{a}(B^{+} \to \pi^{0} K^{+} )\vert_{\text{mistag}}=N_{B} \, \mathcal B(B^{+} \to K^{+}a) f_{L} \times \epsilon ~.
    \label{eq:decay yes}
\end{align}
Similarly, the event rate for ALP escaping the detector mimicking a $B^{+}\to K^{+}\nu\bar{\nu}$ is given by 
\begin{align}
    N_{a}(B^{+} \to K^{+}\slashed{E} )=N_{B} \, \mathcal B (B^{+}\to K^{+} a)(1-f_{L}) ~.
    \label{eq:decay no}
\end{align}

\begin{figure}[tb]
    \centering
    \includegraphics[width=0.48\textwidth]{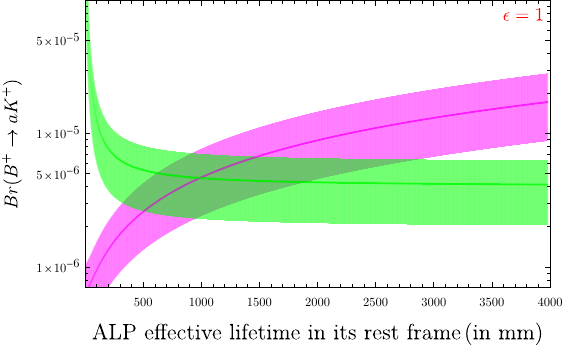} \\[16pt]
    \includegraphics[width=0.48\textwidth]{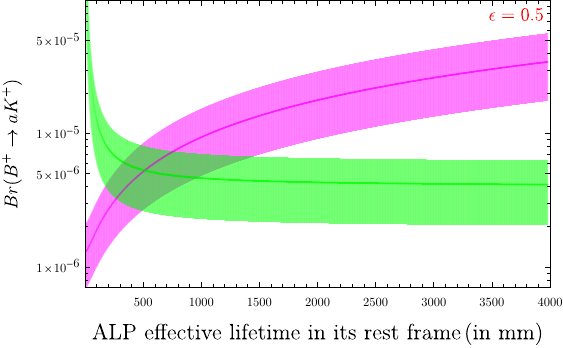} ~~~
    \includegraphics[width=0.48\textwidth]{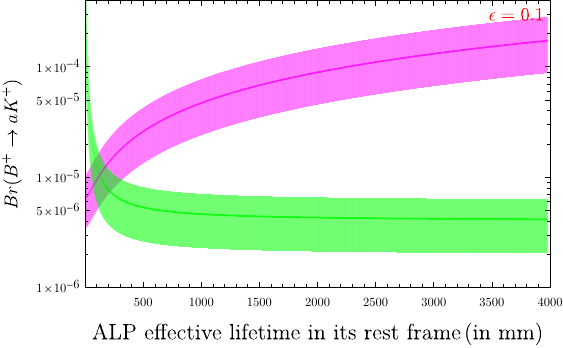}
    \caption{Estimate of the $B^{+} \to a K^{+} $  decay rate as a function of the ALP lifetime. The magenta band corresponds to $B^{+}\to a K^{+} $ decay rate being misidentified as $B^{+} \to \pi^{0} K^{+}$ decay where both $a$ and $\pi^{0}$ decay to two photons. The green band corresponds to the $B^{+} \to K^{+} \nu\bar{\nu}$ decay rate excess above the SM background ($B^{+} \to K^{+} a$ decay where the $a$ escapes the detector). The effect of the detection efficiency of $a\to \gamma\gamma$ decays inside the detector is indicated by the factor $\epsilon$ which is chosen to be $1$, $0.5$ and $0.1$ for the three cases. The bands correspond to $1\sigma$ uncertainties.}
    \label{fig:axion decay length}
\end{figure}


It is likely that not all the $a\to \gamma \gamma$ decays pass the experimental selection criteria to be identified as a $\pi^{0}$. Therefore, we included an additional parameter in the form of detection efficiency $\epsilon$ in Eq.~\eqref{eq:decay yes}. The $B^{+}\to K^{+}\nu\bar{\nu}$ branching fraction measurement on the other hand requires to veto any additional calorimeter activity involving photons~\cite{Belle-II:2023esi}. 

In case of all the $a\to \gamma \gamma$ events identified as $\pi^{0}$ decays (corresponding to $\epsilon=1$),
we infer from the upper plot in Fig.~\ref{fig:axion decay length} that the minimum value for the ALP decay length in its rest frame is around 34 centimeters and extends to several meters. To cover also other options, we  assume two additional benchmark values of the detection efficiency $\epsilon$ to be \textbf{a}) $\epsilon=0.5$ and \textbf{b}) $\epsilon=0.1$ and estimate the ALP decay length in its rest frame. For $\epsilon=0.5$ the minimum and maximum value of $c\tau_{0}$ turn out to be 22 cm and 1.78 m respectively. Similarly we find the minimum and maximum value of $c\tau_{0}$ to be 22 cm and 42 cm respectively in case of $\epsilon$ to be 0.1. 

Moreover, we anticipate that a displaced ALP decaying to two photons can give rise to substantial bias in the measurement of the di-photon invariant mass ($m_{\gamma\gamma}$) and therefore introduce large uncertainty in the estimation of the ALP mass~\cite{BaBar:2021ich}. It is also possible that the two photons emerging from the ALP decay vertex end up as a merged track on the detector leading to rejection of signal events if the angular separation at the electromagnetic calorimeter is less than $3^{\degree}$~\cite{heidelbachmsc:22}. Both of these scenarios would thus require a dedicated analysis with precise modeling of the detector response at Belle II that is not attempted in this work. We encourange a future experimental analysis from Belle II probing the $B^{\pm}\to K^{\pm}a(\to \gamma\gamma)$ signal for an ALP mass around $m_{\pi^{0}}$ as this provides a direct constraints on the $B^{+}\to a K^{+}$ branching fraction in our scenario. In order to find the true $B^{+}\to a K^{+}$ branching fraction we use Eq.~\eqref{eq:decay yes} and Eq.~\eqref{eq:decay no} and get, assuming $\epsilon = 1$
\begin{align}
    \mathcal{B}(B^{+}\to a K^{+})=(4.64_{-2.29}^{+2.40})\times 10^{-6} ~, \qquad \epsilon = 1 ~.
\end{align}
For the lower efficiencies $\epsilon=0.5$ and $\epsilon=0.1$, we find instead
\begin{eqnarray}
    \mathcal{B}(B^{+}\to a K^{+}) &=& (5.29_{-2.60}^{+2.79})\times 10^{-6} ~,~ \qquad \epsilon = 0.5 ~, \\
    \mathcal{B}(B^{+}\to a K^{+}) &=& (10.56_{-5.14}^{+5.83})\times 10^{-6} ~, \qquad \epsilon = 0.1 ~.
\end{eqnarray}
We infer the flavor-changing coupling $h_{sb}^{S}$ introduced in Eq.~\eqref{Eq:ALP lag} in case of $\epsilon=1$ using Eq.~\eqref{Eq:BAK dec rate},
\begin{align}
    h_{sb}^{S}=(10.01_{-2.88}^{+2.32})\times 10^{-9} ~,
\end{align}
and slightly larger values for $\epsilon=1$ and $\epsilon=0.1$. 
The flavor-changing coupling $h_{sb}^{S}$ easily satisfies the constraints from $\Delta F=2$ observable~\cite{Gavela:2019wzg} as well as constraints from previous $B\to K\,+\, \text{missing energy}$ experimental searches~\cite{BaBar:2013npw,BaBar:2021ich}. The ALP lifetime that we find can be readily translated into the ALP decay rate and the  ALP-photon coupling using Eq.~\eqref{eq:G_gg}.

\begin{figure}[tb]
    \centering
    \includegraphics[width=0.45\linewidth]{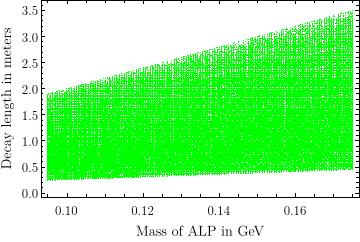}
    \includegraphics[width=0.50\linewidth]{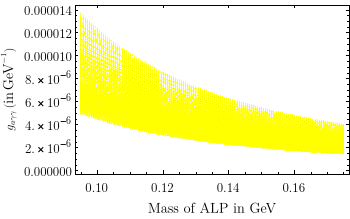}
    \caption{Left: the decay length ($0.25\text{m} < l < 3.5\text{m}$) of the ALP for a range of mass values around the pion mass. Right: the corresponding value of the ALP-photon coupling.}
    \label{fig:massvslength}
\end{figure}

In Fig.~\ref{fig:massvslength} we show the $1\sigma$ range of the preferred decay length and the corresponding ALP-photon coupling as a function of the ALP mass in the vicinity of the pion mass, assuming $\epsilon = 1$. The decay length is smaller and the ALP-photon coupling is larger for $\epsilon < 1$.

\section{Sensitivity to future beam dump experiments} \label{Sec:IV}

\begin{figure}
    \centering
         \includegraphics[width=0.7\linewidth]{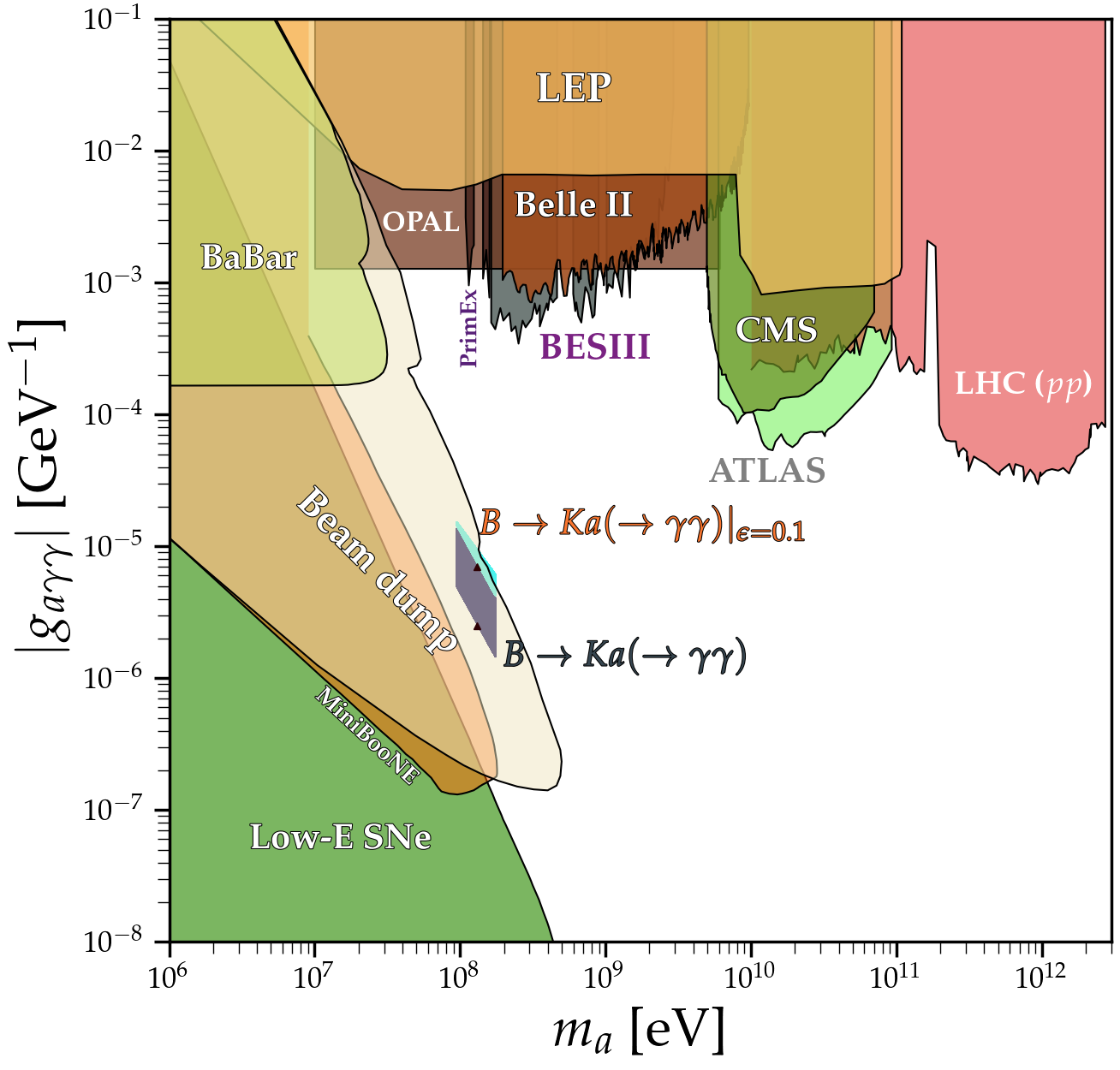}
        \caption{Bounds on the $g_{a\gamma\gamma}-m_{a}$ parameter space presented using \texttt{AxionLimits}~\cite{AxionLimits}. The purple ($\epsilon = 1$) and light blue ($\epsilon = 0.1$) regions corresponds to the preferred $1\sigma$ region based on an ALP having a mass and a $g_{a\gamma\gamma}$ coupling consistent with a joint explanation of $B\to K\nu\bar{\nu}$ excess and $B\to \pi K$-puzzle. The two black triangles indicate two benchmark points that are discussed in more detail in the text.}
    \label{fig:axiolimits}
\end{figure}

In this section, we turn our attention towards the preferred parameter space for the ALP-photon coupling $g_{a\gamma\gamma}$ for ALP masses in the MeV-GeV range and discuss existing constraints from previous beam dump experiments~\cite{CHARM:1985nku,CHARM:1985anb,Blumlein:1990ay,Duffy:1988rw} shown in Fig.~\ref{fig:axiolimits}, as well as the prospects to probe the parameter space~\cite{Ovchynnikov:2023cry,Jerhot:2022chi} at future experiments. 

Interestingly, a number of upcoming experiments like LUXE~\cite{Abramowicz:2021zja,Bai:2021gbm}, SHiP~\cite{Alekhin:2015byh}, FASER~\cite{Feng:2018pew}, DarkQuest~\cite{Blinov:2021say} are sensitive to a region of parameter space with an ALP mass close to the pion mass and an ALP-photon coupling preferred by the joint explanation of the $B\to \pi K$ puzzle and $B\to K\nu\bar{\nu}$ excess events that we identified. 

It is worth emphasizing that the constraints shown in Fig.~\ref{fig:axiolimits} only take into account the ALP-photon coupling ($g_{a\gamma\gamma}$) independently of the additional effective flavor-changing $b-s-a$ coupling considered in our work. In the presence of the flavor-changing coupling, complementary constraints are provided by ongoing experimental efforts at the $B$ factories. In particular, the search in~\cite{BaBar:2021ich} constrains additional parameter space for ALP masses above the pion mass. At beam dump experiments, additional ALP production modes from $B$ meson decays open up, as we now discuss. 

With a proton beam energy of $400$~GeV, both the CHARM experiment and the planned SHiP experiment have a sufficient center of mass energy to produce energetic $b$ hadrons. We will focus on these two experiments in the following.
In fact, the production of $b$ hadrons in previous proton beam dump experiments are known to be moderate~\cite{Lourenco:2006vw, Clarke:2013aya} and SHiP~\cite{Alekhin:2015byh} with its higher total number of proton-nucleus collisions is ideally suited to detect a significant number of long-lived ALPs produced in $B$ meson decays. 

To estimate the total number of $B$ mesons that will be produced at SHiP, we follow~\cite{SHiP:2018xqw}
\begin{align}
    N_{B}=N_{\text{PoT}} \times X_{\bar b b} \times f_\text{cascade} \times ( f_{B^+} + f_{B^0})  \simeq 9.1 \times 10^{13} ~,
\end{align}
where $N_{\text{PoT}} \simeq 2\times 10^{20}$ is the number of protons on target, $X_{\bar b b} \simeq 1.6 \times 10^{-7}$ is the beauty production fraction, $f_\text{cascade} \simeq 1.7$ a cascade enhancement factor, and $f_{B^+} \simeq 0.417$ and $f_{B^0} \simeq 0.418$ are the $B$ meson hadronization fractions.

We estimate the number of detectable $a\to \gamma\gamma$ events (assuming no SM background) based on the experimental setup proposed in the SPS ECN3 high-intensity beam facility with~\cite{SHiP:2023,SHiP:ECN3}. We use the differential distribution of $B^{+}$ production by a proton beam hitting a Molybdenum target from~\cite{Ruf:2115534,Ovchynnikov:2023cry} and calculate the number of events intersecting the surface area of the detector assuming ideal reconstruction efficiency
\begin{align}
    N_{\gamma\gamma}^{\text{obs}}=N_{\text{PoT}}\int \frac{d\sigma(pN\to B)}{dE_{B}d\theta_{B}}\mathcal{B}(B\to aK)\varepsilon_{\text{det}}(z,\theta_{a})\frac{dP_{\text{dec}}}{dz}dz dE_{B}d\theta_{B}
\end{align}
where $P_{\text{dec}}=\exp{(-\frac{\Delta l(z,\theta_{a})}{\gamma\beta c\tau_{0}})}$, $\varepsilon_{\text{det}}$ is the geometrical acceptance for the di-photon signal and $\gamma\beta$ is the boost factor for the ALP in the laboratory frame. The probability of the $a\to \gamma \gamma$ decay happening within the detector volume at a distance $\Delta l$ from the point of production is given by the last factor (assuming a detection efficiency of $\epsilon=1$). We have used a Monte Carlo simulation to calculate the number of detectable events in two benchmark scenarios indicated by black triangles in the plot of Fig.~\ref{fig:axiolimits}
\begin{itemize}
\item \textbf{Scenario (a)} an ALP decay length at the upper $1\sigma$ boundary of the preferred range for $\epsilon = 1$, $c\tau_{0}=2.67\,$m ($g_{a\gamma\gamma}=2.51\times 10^{-6} \,\text{GeV}^{-1}$) corresponding to a branching ratio $\mathcal{B}(B^{+}\to a K^{+})=6.34\times 10^{-6}$;
\item \textbf{Scenario (b)} an ALP decay length at the lower $1\sigma$ boundary of the preferred range for $\epsilon = 1$, $c\tau_{0}=0.34\,$m ($g_{a\gamma\gamma}=7.02\times 10^{-6}\, \text{GeV}^{-1}$) corresponding to a branching ratio of $\mathcal{B}(B^{+}\to a K^{+})=3.04\times 10^{-6}$, respectively. 
\end{itemize}
We find that the number of detectable $a\to \gamma \gamma$ produced in $B\to a K$ decay in \textbf{Scenario (a)} is $N_{\gamma\gamma}^{\text{obs}}\approx 1.14\times 10^{6}$ and in \textbf{Scenario (b)} is $N_{\gamma\gamma}^{\text{obs}}\approx 4.93\times 10^{6}$ respectively. Such large event numbers would unambiguously point to new physics, even in the presence of SM backgrounds. 

Given the large numbers of expected events at SHiP, we also attempted an estimate of the number of $B\to a(\to\gamma\gamma) K$ events passing the selection criteria in the CHARM experiment. Based on the luminosity, location of the decay volume and mean geometric acceptance of the CHARM experiment~\cite{CHARM:1985anb,CHARM:1985nku} our estimate of the number of events for \textbf{Scenario (a)} is $N_{\gamma \gamma}^{\text{obs}}=68$ and for \textbf{Scenario (b)} is $N_{\gamma \gamma}^{\text{obs}}=99$ events assuming ideal detector efficiency. We note that these numbers come with $\mathcal O(1)$ uncertainties in part due to the limited sample size of our Monte Carlo run and in part due to partially known event selection criteria at CHARM. The considered benchmark points may very well be allowed by the non-observation of a long-lived ALP at CHARM.

It is also important to note that for a fixed value of the ALP lifetime $c\tau_{0}$, the ALP-photon coupling is correlated with the mass of the ALP which is chosen to be the $\pi^{0}$ mass in both of the benchmark cases. In reality, as already mentioned above, the uncertainty in the inferred ALP mass is likely non-negligible in the case that the decay to two photons happens significantly away from the point of the ALP production. For ALP masses slightly away from $\pi^{0}$, the preferred parameter space for the ALP-photon coupling opens up further, as we illustrate in Fig~.\ref{fig:axiolimits}. Also note that the shown regions for $\epsilon = 1$ and $\epsilon = 0.1$ correspond to the $1\sigma$ regions. It is likely that part of the preferred region predicts event numbers at the CHARM experiment that are consistent with zero. However, we expect that the shown preferred parameter space will be fully explored by the SHiP experiment in the future, given our estimate of $\mathcal O(10^6)$ events at SHiP for the benchmark points.

Moreover, we note that an auxiliary handle to change the lifetime of the ALP is an additional invisible decay width into a dark sector. Switching on an $\mathcal O(1)$ invisible branching ratio will need to be compensated by a larger ALP-photon coupling to produce a sufficient number of events to address the $B \to \pi K$ puzzle. Both the additional invisible decay width and the increased partial decay width to photons reduce the lifetime of the ALP. In that way, one can arbitrarily suppress the event numbers in the CHARM experiment (with its decay volume $\sim 500$\,m away from the target). We found no events at CHARM for a lifetime of $c\tau_{0} < 0.1\,$m. The suppression of events is much less pronounced for SHiP as its decay volume is much closer to the target.

Finally, we should note that the presence of the ALP-photon coupling by itself induces photo-production of the ALP through Primakoff conversion independently of $B$ decays~\cite{Budnev:1975poe,Dobrich:2015jyk,Feng:2018pew,Aloni:2019ruo,Harland-Lang:2019zur,Dobrich:2019dxc,Dusaev:2020gxi,Balkin:2023gya,Wu:2024fsf}.  (See Appendix~\eqref{App:2} for details of our implementation of the Primakoff production. )

Once again, using Monte Carlo simulations we estimate the number of signal candidates at SHiP passing the experimental cuts for the two above mentioned scenarios and find $N_{\gamma\gamma}^{\text{obs}} \approx 70 $ for \textbf{Scenario (a)} and $N_{\gamma\gamma}^{\text{obs}} \approx 695$ for \textbf{Scenario (b)} respectively. Clearly in both scenarios, the number of ALPs produced through the Primakoff process and decaying to $a\to \gamma \gamma$ is orders of magnitude smaller than the number one can expect from the $B$ decays. To a very good approximation, the sensitivity of SHiP to the ALPs in our scenario is determined by the ALP production through $B$ decays. We expect the same conclusion to hold also for the CHARM experiment.

\section{Conclusion} \label{Sec:V}

In this paper, we have discussed the production of an ALP in $B$ decays that can account for the $B^{+}\to K^{+}\nu\bar{\nu}$ excess events and at the same time mitigate the longstanding $B\to \pi K$ puzzle. The main idea is that an ALP with a lifetime of the order of meters will decay occasionally inside and sometimes outside a detector like Belle II. If the ALP decays outside the detector, it mimics the $B^{+}\to K^{+}\nu\bar{\nu}$ decay signature. If it decays inside the detector into two photons and if it has a mass close to the pion mass, it might get reconstructed as a neutral pion and thus mimic the $B\to \pi^0 K$ decay. 

We find that a $B \to a K$ branching fraction of the order of $3.0 \times 10^{-6}$ - $6 \times 10^{-6}$ and an ALP lifetime ranging from $\sim 0.3$\,m - $3$\,m provides a consistent explanation of both the $B^{+}\to K^{+}\nu\bar{\nu}$ excess events and the $B\to \pi K$ puzzle, if ALP decays to two photons are reconstructed as neutral pions with an efficiency of $\epsilon = 1$. It is conceivable that an ALP that decays to two photons inside the detector but significantly displaced from the $B$ decay vertex might not be reconstructed as a neutral pion. We find that for a smaller efficiency of $\epsilon = 0.1$, the preferred lifetime is only slightly shorter, around $\sim 0.2$\,m.
A more detailed study of the reconstruction efficiency would be desirable and is left for future work.

We considered a minimal ALP setup, taking into account the only two necessary couplings of the ALP, namely a flavor-changing coupling to bottom and strange quarks (which sets the $B \to a K$ branching fraction) and the coupling to photons (which sets the ALP lifetime). Interestingly, the preferred coupling to photons, $g_{a\gamma\gamma} \simeq 10^{-6}\,\text{GeV}^{-1}$ - $10^{-5}\,\text{GeV}^{-1}$ is in a region of parameter space that can be probed by beam dump experiments. We find that the CHARM experiment might have had sensitivity to the interesting parameter space and we estimate that the planned SHiP experiment might produce millions of ALPs that decay to photons in the SHiP detector.

In non-minimal setups, the interesting parameter space opens up considerably. In particular, if the ALP has an additional invisible decay width, larger couplings to photons are required to provide sufficient events that mimic $B \to \pi^0 K$ decays. With a larger coupling to photons, the ALP decay length decreases, requiring a shorter baseline distance for effective detection at beam dump experiments. A systematic study of such non-minimal setups will be presented elsewhere.

\section*{Acknowledgment}

SR is indebted to Ulrik Egede, Rueven Balkin, Stefania Gori for discussions that helped improving the manuscript. SR also acknowledges valuable communications with Ciaran O'Hare and Jan Jerhot regarding clarifications on the limits derived from the beam dump experiments. SR is grateful to the organizers of BEACH 2024 for the opportunity to present part of this work and for the insightful feedback from the participants. Additionally, SR would like to thank the Santa Cruz Institute for Particle Physics (SCIPP) and its members for their hospitality during the completion of this work. During the initial phase of the project, SR was financially supported by a postdoctoral fellowship from Chennai Mathematical Institute.  The research of WA is supported by the U.S. Department of Energy grant number DE-SC0010107.

\begin{appendix}
\section{ALP Decay Probability} \label{App:1}

In this appendix we provide a simple analytical approximation for the probability that an ALP produced from $B$ mesons decays within a distance $l_\text{max}$ from the production point, see Eq.~\ref{eq: survival probability int}. Setting the production angle ($\theta$) of the $B$ mesons close to 0 we find
\begin{align}
    f_{L}=1-\int_{0}^{\pi/2} \sin{\theta_{a}} d\theta_{a} \exp{\left(-\frac{m_{a}l_{max}}{c\tau_{0}\gamma_{B}(p_{a}\cos{\theta_{a}}+\beta_{B} E_{a})}\right)} ~.
\end{align}
This integral can be recast in the form
\begin{align}
    \int_{z_{1}}^{z_{2}} dz\,  \frac{\exp(-Az)}{z^{2}}=\frac{\exp(-Az_{1})}{z_{1}}-\frac{\exp(-Az_{2})}{z_{2}}-A \big(Ei(-Az_{2})-Ei(-Az_{1}) \big) ~,
\end{align}
where $Ei(-z)=-E_{1}(z)$, and the definition of exponential integral is
\begin{align}
    E_{n}(z)=\int_{1}^{\infty}\exp{(-zt)}\,t^{-n}dt ~,\qquad n=0,1,\cdots,  \qquad \text{Re}(z) > 0 ~.
\end{align}

\section{Primakoff production of the ALP at proton beam-dump experiments} \label{App:2}

Following~\cite{Dobrich:2015jyk,Wu:2024fsf} we calculate the differential cross-section for $\gamma + N\to a + N$~\cite{Feng:2018pew} 
\begin{align}
    \frac{d\sigma(\gamma \, N \to a N)}{d\theta_{a}}=\frac{\alpha g_{\alpha \gamma \gamma}^{2} Z^{2} F(\vert t \vert)^{2}}{4}\frac{k_{a}^{4}\sin{\theta_{a}}^{3}}{t^{2}} ~,
\end{align}
and use the equivalent photon approximation~\cite{Liu:2016mqv,Liu:2017htz} to obtain total $p+N\to p+N+a$ cross section.
The 4-vectors in laboratory frame for the almost on-shell incoming photon, the nucleus at rest, the outgoing ALP, and the scattered nucleus are respectively, 
\begin{align}
p^{\mu}=
    \begin{pmatrix}
        E_{\gamma}\\
        p_{t}\\
        0\\
        E_{\gamma}
    \end{pmatrix}
    ,\quad
P_{i}^{\mu}=     \begin{pmatrix}
        m_{N}\\
        0\\
        0\\
        0
    \end{pmatrix}
    ,\quad
k^{\mu}= \begin{pmatrix}
        E_{a}\\
        k_{a}\sin{\theta_{a}}\\
        0\\
        k_{a}\cos{\theta_{a}}
    \end{pmatrix}
    ,\quad    
P_{f}^{\mu}=p^{\mu}+P^{\mu}_{i}-k^{\mu}   ~. 
\end{align}
Here $m_{a},\,m_{N}$ are the mass of the ALP and the mass of the scattering nucleus, $E_{\gamma},\, E_{a}$ the incoming photon and outgoing ALP energy in the laboratory frame, $p_{t}$ is the transverse photon momentum, $\theta_{a}$ the angle between the ALP momentum and the beam direction, and $E_{\gamma}\simeq E_{a}$ for $\theta_{a}\ll 1$ and $m_{a},p_{t}\ll E_{a},m_{N}$. In the case of elastic scattering, $P_{i}^{2}=P_{f}^{2}=m_{N}^{2}$ and the complete expression for $E_{\gamma}$ is given by, 
\begin{align}
    E_{\gamma}=\frac{m_{N}(k_{a}^{2}+m_{a}^{2})^{1/2}-\frac{m_{a}^{2}-p_{t}^{2}+2k_{a}p_{t}\sin{\theta_{a}}}{2}}{m_{N}-(k_{a}^{2}+m_{a}^{2})^{1/2}+k_{a}\cos{\theta_{a}}} ~.
\end{align}
For $\frac{m_{a}}{\sqrt{m_{N}(E_{\gamma}+m_{N})}},\, \theta_{a} \ll 1 $ we can write down an approximate expression for $k_{a}$ as,
\begin{align}
k_{a}\simeq E_{\gamma}\Big(1-\frac{m_{a}^2}{2E_{\gamma}^2}-\frac{p_{t}^2}{2E_{\gamma}m_{N}}\Big)-\frac{m_{a}^{4}}{8E_{\gamma}m_{N}^{2}}-\frac{E_{\gamma}\theta_{a}^{2}}{2m_{N}}\Big(E_{\gamma}-\frac{m_{a}^{2}}{2m_{N}}\Big) ~,
\end{align}
and the momentum transfer $t$,
\begin{align}
t=(P_{i}^{\mu}-P_{f}^{\mu})^{2}=(k^{\mu}-p^{\mu})^{2}\simeq-\frac{m_{a}^{4}}{4E_{\gamma}^{2}}-E_{\gamma}^{2}\theta^{2}-p_{t}^{2}+2E_{\gamma}p_{t}\theta_{a} ~,
\end{align}
is suppressed for a limiting value $t> t_{0}$ in case of elastic scattering. The numerical value of $t_{0}$ is inferred from the vanishing of the charge form factor $F(q^{2}=\vert t_{0}\vert)$ where $F(q^2)$ is given by, 
\begin{align}
    F(q^{2})=\frac{3j_{1}(\sqrt{q^{2}}R_{1})}{\sqrt{q^{2}}R_{1}} \exp{\Big[-\frac{(\sqrt{q^{2}}\textrm{s})^{2}}{2}\Big]}
\end{align} ~,
where $j_{1}$ is the first spherical Bessel function of the first kind and
\begin{align}
    R_{1}=\sqrt{(1.23\,{\rm fm}\,A^{1/3}-0.6\,{\rm fm})^{2}+7\pi^{2}(0.52\,{\rm fm})^{2}/3-5\textrm{s}^{2}} ~, \quad \textrm{s}=0.9\, \text{fm} ~.
\end{align}
To obtain the total cross section for the photoproduction of the ALPs in a proton beam dump experiment, we use the equivalent photon spectrum of the proton beam in the laboratory frame and estimate the distribution of photon momenta $q$ by $\gamma_{p}(x=\frac{E_{\gamma}}{E_{\text{beam}}},p_{t}^{2})$~\cite{Budnev:1975poe,Vysotskii:2018eic,Ma:2021lgv},
\begin{align}
    \gamma(x,p_{t}^{2})=\frac{\alpha}{\pi^{2}x}\frac{1}{(p_{t}^{2}+x^{2}m_{p}^2)}\Big[\frac{p_{t}^2}{(p_{t}^{2}+x^{2}m_{p}^2)}(1-x)D(q^{2})+\frac{x^2}{2}C(q^{2})\Big] ~,
\end{align}
where for protons,
\begin{align}
    D(q^2)=\frac{4m_{p}^{2}G_{E}^{2}(q^2)+q^{2}G_{M}^{2}(q^2)}{4m_{p}^{2}+q^{2}} ~, \quad C(q^2)=G_{M}^{2}(q^2) ~.
\end{align}
The form factors $G_{E}(q^2)$ and $G_{M}(q^2)$ are set to vanish for $q^{2}> 1\,\text{GeV}$ and are expressed as  
\begin{align}
    G_{E}(q^2)=\frac{1}{(1+q^{2}/q^{2}_{0})^{2}} ~,\qquad   G_{M}(q^2)=\frac{\mu_{p}}{(1+q^{2}/q^{2}_{0})^{2}} ~,
\end{align}
where $q_{0}^{2}=0.71\,\text{GeV}^{2}$ and $\mu_{p}^{2}=7.78$ is the magnetic moment of the proton. Finally the total cross section can be expressed as,
\begin{align}
    \sigma_{\text{Primakoff}}=\frac{1}{ E_{\text{beam}}\sigma_{pN}}\int  \gamma_{p}(E_{a}/E_{\text{beam}},p_{t}^{2})\frac{d\sigma(\gamma \, N \to a N)}{d\theta_{a}} d\theta_{a}dE_{a}dp_{t}^{2} ~,
\end{align}
where $\sigma_{pN}=53\text{mb}\times A^{0.77}$. The expected number of detectable ALP in its decay to two photons through Primakoff production is obtained by plugging in the proton beam energy ($E_{\text{beam}}=400\,\text{GeV}$), the atomic number and mass of the target nucleus ($Z=42,\, A=96$) in the differential distribution and weighing in the geometry factor that ensures that ALP and its decay products both remain within the detector volume. 
\end{appendix}

\bibliographystyle{JHEP}
\bibliography{btokinv}

\providecommand{\href}[2]{#2}\begingroup\raggedright\begin{thebibliography}{100}

\bibitem{Kobayashi:1973fv}
M.~Kobayashi and T.~Maskawa, \emph{{CP Violation in the Renormalizable Theory
  of Weak Interaction}}, \href{https://doi.org/10.1143/PTP.49.652}{\emph{Prog.
  Theor. Phys.} {\bfseries 49} (1973) 652}.

\bibitem{BaBar:1998yfb}
{\scshape BaBar} collaboration, \emph{{The BABAR physics book: Physics at an
  asymmetric $B$ factory}} (10, 1998),
  \href{https://doi.org/10.2172/979931}{10.2172/979931}.

\bibitem{BaBar:2014omp}
{\scshape BaBar, Belle} collaboration, \emph{{The Physics of the B Factories}},
  \href{https://doi.org/10.1140/epjc/s10052-014-3026-9}{\emph{Eur. Phys. J. C}
  {\bfseries 74} (2014) 3026}
  [\href{https://arxiv.org/abs/1406.6311}{{\ttfamily 1406.6311}}].

\bibitem{Belle:2000cnh}
{\scshape Belle} collaboration, \emph{{The Belle Detector}},
  \href{https://doi.org/10.1016/S0168-9002(01)02013-7}{\emph{Nucl. Instrum.
  Meth. A} {\bfseries 479} (2002) 117}.

\bibitem{LHCb:2012myk}
{\scshape LHCb} collaboration, \emph{{Implications of LHCb measurements and
  future prospects}},
  \href{https://doi.org/10.1140/epjc/s10052-013-2373-2}{\emph{Eur. Phys. J. C}
  {\bfseries 73} (2013) 2373}
  [\href{https://arxiv.org/abs/1208.3355}{{\ttfamily 1208.3355}}].

\bibitem{Belle-II:2018jsg}
{\scshape Belle-II} collaboration, \emph{{The Belle II Physics Book}},
  \href{https://doi.org/10.1093/ptep/ptz106}{\emph{PTEP} {\bfseries 2019}
  (2019) 123C01} [\href{https://arxiv.org/abs/1808.10567}{{\ttfamily
  1808.10567}}].

\bibitem{HFLAV:2022esi}
{\scshape HFLAV} collaboration, \emph{{Averages of b-hadron, c-hadron, and
  \ensuremath{\tau}-lepton properties as of 2021}},
  \href{https://doi.org/10.1103/PhysRevD.107.052008}{\emph{Phys. Rev. D}
  {\bfseries 107} (2023) 052008}
  [\href{https://arxiv.org/abs/2206.07501}{{\ttfamily 2206.07501}}].

\bibitem{ParticleDataGroup:2024cfk}
{\scshape Particle Data Group} collaboration, \emph{{Review of particle
  physics}}, \href{https://doi.org/10.1103/PhysRevD.110.030001}{\emph{Phys.
  Rev. D} {\bfseries 110} (2024) 030001}.

\bibitem{Capdevila:2023yhq}
B.~Capdevila, A.~Crivellin and J.~Matias, \emph{{Review of semileptonic B
  anomalies}},
  \href{https://doi.org/10.1140/epjs/s11734-023-01012-2}{\emph{Eur. Phys. J.
  ST} {\bfseries 1} (2023) 20}
  [\href{https://arxiv.org/abs/2309.01311}{{\ttfamily 2309.01311}}].

\bibitem{Batell:2009jf}
B.~Batell, M.~Pospelov and A.~Ritz, \emph{{Multi-lepton Signatures of a Hidden
  Sector in Rare B Decays}},
  \href{https://doi.org/10.1103/PhysRevD.83.054005}{\emph{Phys. Rev. D}
  {\bfseries 83} (2011) 054005}
  [\href{https://arxiv.org/abs/0911.4938}{{\ttfamily 0911.4938}}].

\bibitem{Freytsis:2009ct}
M.~Freytsis, Z.~Ligeti and J.~Thaler, \emph{{Constraining the Axion Portal with
  $B \to K l^+ l^-$}},
  \href{https://doi.org/10.1103/PhysRevD.81.034001}{\emph{Phys. Rev. D}
  {\bfseries 81} (2010) 034001}
  [\href{https://arxiv.org/abs/0911.5355}{{\ttfamily 0911.5355}}].

\bibitem{Schmidt-Hoberg:2013hba}
K.~Schmidt-Hoberg, F.~Staub and M.W.~Winkler, \emph{{Constraints on light
  mediators: confronting dark matter searches with B physics}},
  \href{https://doi.org/10.1016/j.physletb.2013.11.015}{\emph{Phys. Lett. B}
  {\bfseries 727} (2013) 506}
  [\href{https://arxiv.org/abs/1310.6752}{{\ttfamily 1310.6752}}].

\bibitem{Dolan:2014ska}
M.J.~Dolan, F.~Kahlhoefer, C.~McCabe and K.~Schmidt-Hoberg, \emph{{A taste of
  dark matter: Flavour constraints on pseudoscalar mediators}},
  \href{https://doi.org/10.1007/JHEP03(2015)171}{\emph{JHEP} {\bfseries 03}
  (2015) 171} [\href{https://arxiv.org/abs/1412.5174}{{\ttfamily 1412.5174}}].

\bibitem{Izaguirre:2016dfi}
E.~Izaguirre, T.~Lin and B.~Shuve, \emph{{Searching for Axionlike Particles in
  Flavor-Changing Neutral Current Processes}},
  \href{https://doi.org/10.1103/PhysRevLett.118.111802}{\emph{Phys. Rev. Lett.}
  {\bfseries 118} (2017) 111802}
  [\href{https://arxiv.org/abs/1611.09355}{{\ttfamily 1611.09355}}].

\bibitem{Sala:2017ihs}
F.~Sala and D.M.~Straub, \emph{{A New Light Particle in B Decays?}},
  \href{https://doi.org/10.1016/j.physletb.2017.09.072}{\emph{Phys. Lett. B}
  {\bfseries 774} (2017) 205}
  [\href{https://arxiv.org/abs/1704.06188}{{\ttfamily 1704.06188}}].

\bibitem{Datta:2017ezo}
A.~Datta, J.~Kumar, J.~Liao and D.~Marfatia, \emph{{New light mediators for the
  $R_K$ and $R_{K^*}$ puzzles}},
  \href{https://doi.org/10.1103/PhysRevD.97.115038}{\emph{Phys. Rev. D}
  {\bfseries 97} (2018) 115038}
  [\href{https://arxiv.org/abs/1705.08423}{{\ttfamily 1705.08423}}].

\bibitem{Dror:2017ehi}
J.A.~Dror, R.~Lasenby and M.~Pospelov, \emph{{New constraints on light vectors
  coupled to anomalous currents}},
  \href{https://doi.org/10.1103/PhysRevLett.119.141803}{\emph{Phys. Rev. Lett.}
  {\bfseries 119} (2017) 141803}
  [\href{https://arxiv.org/abs/1705.06726}{{\ttfamily 1705.06726}}].

\bibitem{Dror:2017nsg}
J.A.~Dror, R.~Lasenby and M.~Pospelov, \emph{{Dark forces coupled to
  nonconserved currents}},
  \href{https://doi.org/10.1103/PhysRevD.96.075036}{\emph{Phys. Rev. D}
  {\bfseries 96} (2017) 075036}
  [\href{https://arxiv.org/abs/1707.01503}{{\ttfamily 1707.01503}}].

\bibitem{Altmannshofer:2017bsz}
W.~Altmannshofer, M.J.~Baker, S.~Gori, R.~Harnik, M.~Pospelov, E.~Stamou
  et~al., \emph{{Light resonances and the low-q$^{2}$ bin of $ {R}_{K^{*}} $}},
  \href{https://doi.org/10.1007/JHEP03(2018)188}{\emph{JHEP} {\bfseries 03}
  (2018) 188} [\href{https://arxiv.org/abs/1711.07494}{{\ttfamily
  1711.07494}}].

\bibitem{Winkler:2018qyg}
M.W.~Winkler, \emph{{Decay and detection of a light scalar boson mixing with
  the Higgs boson}},
  \href{https://doi.org/10.1103/PhysRevD.99.015018}{\emph{Phys. Rev. D}
  {\bfseries 99} (2019) 015018}
  [\href{https://arxiv.org/abs/1809.01876}{{\ttfamily 1809.01876}}].

\bibitem{Dobrich:2018jyi}
B.~D\"obrich, F.~Ertas, F.~Kahlhoefer and T.~Spadaro, \emph{{Model-independent
  bounds on light pseudoscalars from rare B-meson decays}},
  \href{https://doi.org/10.1016/j.physletb.2019.01.064}{\emph{Phys. Lett. B}
  {\bfseries 790} (2019) 537}
  [\href{https://arxiv.org/abs/1810.11336}{{\ttfamily 1810.11336}}].

\bibitem{Aloni:2018vki}
D.~Aloni, Y.~Soreq and M.~Williams, \emph{{Coupling QCD-Scale Axionlike
  Particles to Gluons}},
  \href{https://doi.org/10.1103/PhysRevLett.123.031803}{\emph{Phys. Rev. Lett.}
  {\bfseries 123} (2019) 031803}
  [\href{https://arxiv.org/abs/1811.03474}{{\ttfamily 1811.03474}}].

\bibitem{Gavela:2019wzg}
M.B.~Gavela, R.~Houtz, P.~Quilez, R.~Del~Rey and O.~Sumensari, \emph{{Flavor
  constraints on electroweak ALP couplings}},
  \href{https://doi.org/10.1140/epjc/s10052-019-6889-y}{\emph{Eur. Phys. J. C}
  {\bfseries 79} (2019) 369}
  [\href{https://arxiv.org/abs/1901.02031}{{\ttfamily 1901.02031}}].

\bibitem{Filimonova:2019tuy}
A.~Filimonova, R.~Sch\"afer and S.~Westhoff, \emph{{Probing dark sectors with
  long-lived particles at BELLE II}},
  \href{https://doi.org/10.1103/PhysRevD.101.095006}{\emph{Phys. Rev. D}
  {\bfseries 101} (2020) 095006}
  [\href{https://arxiv.org/abs/1911.03490}{{\ttfamily 1911.03490}}].

\bibitem{MartinCamalich:2020dfe}
J.~Martin~Camalich, M.~Pospelov, P.N.H.~Vuong, R.~Ziegler and J.~Zupan,
  \emph{{Quark Flavor Phenomenology of the QCD Axion}},
  \href{https://doi.org/10.1103/PhysRevD.102.015023}{\emph{Phys. Rev. D}
  {\bfseries 102} (2020) 015023}
  [\href{https://arxiv.org/abs/2002.04623}{{\ttfamily 2002.04623}}].

\bibitem{Kachanovich:2020yhi}
A.~Kachanovich, U.~Nierste and I.~Ni\v{s}and\v{z}i\'c, \emph{{Higgs portal to
  dark matter and $B\to K^{(*)}$ decays}},
  \href{https://doi.org/10.1140/epjc/s10052-020-8240-z}{\emph{Eur. Phys. J. C}
  {\bfseries 80} (2020) 669}
  [\href{https://arxiv.org/abs/2003.01788}{{\ttfamily 2003.01788}}].

\bibitem{Chakraborty:2021wda}
S.~Chakraborty, M.~Kraus, V.~Loladze, T.~Okui and K.~Tobioka, \emph{{Heavy QCD
  axion in b\textrightarrow{}s transition: Enhanced limits and projections}},
  \href{https://doi.org/10.1103/PhysRevD.104.055036}{\emph{Phys. Rev. D}
  {\bfseries 104} (2021) 055036}
  [\href{https://arxiv.org/abs/2102.04474}{{\ttfamily 2102.04474}}].

\bibitem{Dutta:2021afo}
B.~Dutta, S.~Ghosh, P.~Huang and J.~Kumar, \emph{{Explaining
  g\ensuremath{\mu}-2 and RK(*) using the light mediators of U(1)T3R}},
  \href{https://doi.org/10.1103/PhysRevD.105.015011}{\emph{Phys. Rev. D}
  {\bfseries 105} (2022) 015011}
  [\href{https://arxiv.org/abs/2105.07655}{{\ttfamily 2105.07655}}].

\bibitem{Darme:2021qzw}
L.~Darm\'e, M.~Fedele, K.~Kowalska and E.M.~Sessolo, \emph{{Flavour anomalies
  and the muon g \ensuremath{-} 2 from feebly interacting particles}},
  \href{https://doi.org/10.1007/JHEP03(2022)085}{\emph{JHEP} {\bfseries 03}
  (2022) 085} [\href{https://arxiv.org/abs/2106.12582}{{\ttfamily
  2106.12582}}].

\bibitem{Bertholet:2021hjl}
E.~Bertholet, S.~Chakraborty, V.~Loladze, T.~Okui, A.~Soffer and K.~Tobioka,
  \emph{{Heavy QCD axion at Belle II: Displaced and prompt signals}},
  \href{https://doi.org/10.1103/PhysRevD.105.L071701}{\emph{Phys. Rev. D}
  {\bfseries 105} (2022) L071701}
  [\href{https://arxiv.org/abs/2108.10331}{{\ttfamily 2108.10331}}].

\bibitem{Bauer:2021mvw}
M.~Bauer, M.~Neubert, S.~Renner, M.~Schnubel and A.~Thamm, \emph{{Flavor probes
  of axion-like particles}},
  \href{https://doi.org/10.1007/JHEP09(2022)056}{\emph{JHEP} {\bfseries 09}
  (2022) 056} [\href{https://arxiv.org/abs/2110.10698}{{\ttfamily
  2110.10698}}].

\bibitem{Ferber:2022rsf}
T.~Ferber, A.~Filimonova, R.~Sch\"afer and S.~Westhoff, \emph{{Displaced or
  invisible? ALPs from B decays at Belle II}},
  \href{https://doi.org/10.1007/JHEP04(2023)131}{\emph{JHEP} {\bfseries 04}
  (2023) 131} [\href{https://arxiv.org/abs/2201.06580}{{\ttfamily
  2201.06580}}].

\bibitem{Crivellin:2022obd}
A.~Crivellin, C.A.~Manzari, W.~Altmannshofer, G.~Inguglia, P.~Feichtinger and
  J.~Martin~Camalich, \emph{{Toward excluding a light Z' explanation of
  b\textrightarrow{}s\ensuremath{\ell}+\ensuremath{\ell}-}},
  \href{https://doi.org/10.1103/PhysRevD.106.L031703}{\emph{Phys. Rev. D}
  {\bfseries 106} (2022) L031703}
  [\href{https://arxiv.org/abs/2202.12900}{{\ttfamily 2202.12900}}].

\bibitem{Bonilla:2022qgm}
J.~Bonilla, A.~de~Giorgi, B.~Gavela, L.~Merlo and M.~Ramos, \emph{{The cost of
  an ALP solution to the neutral B-anomalies}},
  \href{https://doi.org/10.1007/JHEP02(2023)138}{\emph{JHEP} {\bfseries 02}
  (2023) 138} [\href{https://arxiv.org/abs/2209.11247}{{\ttfamily
  2209.11247}}].

\bibitem{Bonilla:2022vtn}
J.~Bonilla, A.~de~Giorgi and M.~Ramos, \emph{{Neutral $B$-anomalies from an
  $\mathit{on\text{-}shell}$ scalar exchange}},
  \href{https://arxiv.org/abs/2211.05135}{{\ttfamily 2211.05135}}.

\bibitem{He:2022ljo}
X.-G.~He, X.-D.~Ma and G.~Valencia, \emph{{FCNC B and K meson decays with light
  bosonic Dark Matter}},
  \href{https://doi.org/10.1007/JHEP03(2023)037}{\emph{JHEP} {\bfseries 03}
  (2023) 037} [\href{https://arxiv.org/abs/2209.05223}{{\ttfamily
  2209.05223}}].

\bibitem{Zhang:2023wlt}
Y.~Zhang, A.~Ishikawa, E.~Kou, D.T.~Marcantonio and P.~Urquijo, \emph{{Belle II
  observation prospects for axionlike particle production from B meson
  annihilation decay}},
  \href{https://doi.org/10.1103/PhysRevD.109.016008}{\emph{Phys. Rev. D}
  {\bfseries 109} (2024) 016008}
  [\href{https://arxiv.org/abs/2306.03464}{{\ttfamily 2306.03464}}].

\bibitem{Ovchynnikov:2023von}
M.~Ovchynnikov, M.A.~Schmidt and T.~Schwetz, \emph{{Complementarity of
  $B\rightarrow K^{(*)} \mu \bar{\mu }$ and $B\rightarrow K^{(*)} +
  \textrm{inv}$ for searches of GeV-scale Higgs-like scalars}},
  \href{https://doi.org/10.1140/epjc/s10052-023-11975-0}{\emph{Eur. Phys. J. C}
  {\bfseries 83} (2023) 791}
  [\href{https://arxiv.org/abs/2306.09508}{{\ttfamily 2306.09508}}].

\bibitem{Ishida:2020oxl}
H.~Ishida, S.~Matsuzaki and Y.~Shigekami, \emph{{New perspective in searching
  for axionlike particles from flavor physics}},
  \href{https://doi.org/10.1103/PhysRevD.103.095022}{\emph{Phys. Rev. D}
  {\bfseries 103} (2021) 095022}
  [\href{https://arxiv.org/abs/2006.02725}{{\ttfamily 2006.02725}}].

\bibitem{Bhattacharya:2021ggm}
B.~Bhattacharya, A.~Datta, D.~Marfatia, S.~Nandi and J.~Waite,
  \emph{{Axion-like particles resolve the $B \to \pi K$ and $g-2$ anomalies}},
  \href{https://doi.org/10.1103/PhysRevD.104.L051701}{\emph{Phys. Rev. D}
  {\bfseries 104} (2021) L051701}
  [\href{https://arxiv.org/abs/2104.03947}{{\ttfamily 2104.03947}}].

\bibitem{Mimasu:2014nea}
K.~Mimasu and V.~Sanz, \emph{{ALPs at Colliders}},
  \href{https://doi.org/10.1007/JHEP06(2015)173}{\emph{JHEP} {\bfseries 06}
  (2015) 173} [\href{https://arxiv.org/abs/1409.4792}{{\ttfamily 1409.4792}}].

\bibitem{Jaeckel:2015jla}
J.~Jaeckel and M.~Spannowsky, \emph{{Probing MeV to 90 GeV axion-like particles
  with LEP and LHC}},
  \href{https://doi.org/10.1016/j.physletb.2015.12.037}{\emph{Phys. Lett. B}
  {\bfseries 753} (2016) 482}
  [\href{https://arxiv.org/abs/1509.00476}{{\ttfamily 1509.00476}}].

\bibitem{Knapen:2016moh}
S.~Knapen, T.~Lin, H.K.~Lou and T.~Melia, \emph{{Searching for Axionlike
  Particles with Ultraperipheral Heavy-Ion Collisions}},
  \href{https://doi.org/10.1103/PhysRevLett.118.171801}{\emph{Phys. Rev. Lett.}
  {\bfseries 118} (2017) 171801}
  [\href{https://arxiv.org/abs/1607.06083}{{\ttfamily 1607.06083}}].

\bibitem{Alves:2017avw}
D.S.M.~Alves and N.~Weiner, \emph{{A viable QCD axion in the MeV mass range}},
  \href{https://doi.org/10.1007/JHEP07(2018)092}{\emph{JHEP} {\bfseries 07}
  (2018) 092} [\href{https://arxiv.org/abs/1710.03764}{{\ttfamily
  1710.03764}}].

\bibitem{Dolan:2017osp}
M.J.~Dolan, T.~Ferber, C.~Hearty, F.~Kahlhoefer and K.~Schmidt-Hoberg,
  \emph{{Revised constraints and Belle II sensitivity for visible and invisible
  axion-like particles}},
  \href{https://doi.org/10.1007/JHEP12(2017)094}{\emph{JHEP} {\bfseries 12}
  (2017) 094} [\href{https://arxiv.org/abs/1709.00009}{{\ttfamily
  1709.00009}}].

\bibitem{Brivio:2017ije}
I.~Brivio, M.B.~Gavela, L.~Merlo, K.~Mimasu, J.M.~No, R.~del Rey et~al.,
  \emph{{ALPs Effective Field Theory and Collider Signatures}},
  \href{https://doi.org/10.1140/epjc/s10052-017-5111-3}{\emph{Eur. Phys. J. C}
  {\bfseries 77} (2017) 572}
  [\href{https://arxiv.org/abs/1701.05379}{{\ttfamily 1701.05379}}].

\bibitem{Bauer:2017ris}
M.~Bauer, M.~Neubert and A.~Thamm, \emph{{Collider Probes of Axion-Like
  Particles}}, \href{https://doi.org/10.1007/JHEP12(2017)044}{\emph{JHEP}
  {\bfseries 12} (2017) 044}
  [\href{https://arxiv.org/abs/1708.00443}{{\ttfamily 1708.00443}}].

\bibitem{Bjorkeroth:2018dzu}
F.~Bj\"orkeroth, E.J.~Chun and S.F.~King, \emph{{Flavourful Axion
  Phenomenology}}, \href{https://doi.org/10.1007/JHEP08(2018)117}{\emph{JHEP}
  {\bfseries 08} (2018) 117}
  [\href{https://arxiv.org/abs/1806.00660}{{\ttfamily 1806.00660}}].

\bibitem{Bauer:2018uxu}
M.~Bauer, M.~Heiles, M.~Neubert and A.~Thamm, \emph{{Axion-Like Particles at
  Future Colliders}},
  \href{https://doi.org/10.1140/epjc/s10052-019-6587-9}{\emph{Eur. Phys. J. C}
  {\bfseries 79} (2019) 74} [\href{https://arxiv.org/abs/1808.10323}{{\ttfamily
  1808.10323}}].

\bibitem{CidVidal:2018blh}
X.~Cid~Vidal, A.~Mariotti, D.~Redigolo, F.~Sala and K.~Tobioka, \emph{{New
  Axion Searches at Flavor Factories}},
  \href{https://doi.org/10.1007/JHEP01(2019)113}{\emph{JHEP} {\bfseries 01}
  (2019) 113} [\href{https://arxiv.org/abs/1810.09452}{{\ttfamily
  1810.09452}}].

\bibitem{Dobrich:2019dxc}
B.~D\"obrich, J.~Jaeckel and T.~Spadaro, \emph{{Light in the beam dump - ALP
  production from decay photons in proton beam-dumps}},
  \href{https://doi.org/10.1007/JHEP05(2019)213}{\emph{JHEP} {\bfseries 05}
  (2019) 213} [\href{https://arxiv.org/abs/1904.02091}{{\ttfamily
  1904.02091}}].

\bibitem{Merlo:2019anv}
L.~Merlo, F.~Pobbe, S.~Rigolin and O.~Sumensari, \emph{{Revisiting the
  production of ALPs at B-factories}},
  \href{https://doi.org/10.1007/JHEP06(2019)091}{\emph{JHEP} {\bfseries 06}
  (2019) 091} [\href{https://arxiv.org/abs/1905.03259}{{\ttfamily
  1905.03259}}].

\bibitem{Altmannshofer:2019yji}
W.~Altmannshofer, S.~Gori and D.J.~Robinson, \emph{{Constraining axionlike
  particles from rare pion decays}},
  \href{https://doi.org/10.1103/PhysRevD.101.075002}{\emph{Phys. Rev. D}
  {\bfseries 101} (2020) 075002}
  [\href{https://arxiv.org/abs/1909.00005}{{\ttfamily 1909.00005}}].

\bibitem{Hook:2019qoh}
A.~Hook, S.~Kumar, Z.~Liu and R.~Sundrum, \emph{{High Quality QCD Axion and the
  LHC}}, \href{https://doi.org/10.1103/PhysRevLett.124.221801}{\emph{Phys. Rev.
  Lett.} {\bfseries 124} (2020) 221801}
  [\href{https://arxiv.org/abs/1911.12364}{{\ttfamily 1911.12364}}].

\bibitem{Gori:2020xvq}
S.~Gori, G.~Perez and K.~Tobioka, \emph{{KOTO vs. NA62 Dark Scalar Searches}},
  \href{https://doi.org/10.1007/JHEP08(2020)110}{\emph{JHEP} {\bfseries 08}
  (2020) 110} [\href{https://arxiv.org/abs/2005.05170}{{\ttfamily
  2005.05170}}].

\bibitem{Blinov:2021say}
N.~Blinov, E.~Kowalczyk and M.~Wynne, \emph{{Axion-like particle searches at
  DarkQuest}}, \href{https://doi.org/10.1007/JHEP02(2022)036}{\emph{JHEP}
  {\bfseries 02} (2022) 036}
  [\href{https://arxiv.org/abs/2112.09814}{{\ttfamily 2112.09814}}].

\bibitem{Dreyer:2021aqd}
S.~Dreyer et~al., \emph{{Physics reach of a long-lived particle detector at
  Belle II}},  \href{https://arxiv.org/abs/2105.12962}{{\ttfamily 2105.12962}}.

\bibitem{Bandyopadhyay:2021wbb}
T.~Bandyopadhyay, S.~Ghosh and T.S.~Roy, \emph{{ALP-Pions generalized}},
  \href{https://doi.org/10.1103/PhysRevD.105.115039}{\emph{Phys. Rev. D}
  {\bfseries 105} (2022) 115039}
  [\href{https://arxiv.org/abs/2112.13147}{{\ttfamily 2112.13147}}].

\bibitem{Jerhot:2022chi}
J.~Jerhot, B.~D\"obrich, F.~Ertas, F.~Kahlhoefer and T.~Spadaro,
  \emph{{ALPINIST: Axion-Like Particles In Numerous Interactions Simulated and
  Tabulated}}, \href{https://doi.org/10.1007/JHEP07(2022)094}{\emph{JHEP}
  {\bfseries 07} (2022) 094}
  [\href{https://arxiv.org/abs/2201.05170}{{\ttfamily 2201.05170}}].

\bibitem{Altmannshofer:2022ckw}
W.~Altmannshofer, J.A.~Dror and S.~Gori, \emph{{New Opportunities for Detecting
  Axion-Lepton Interactions}},
  \href{https://doi.org/10.1103/PhysRevLett.130.241801}{\emph{Phys. Rev. Lett.}
  {\bfseries 130} (2023) 241801}
  [\href{https://arxiv.org/abs/2209.00665}{{\ttfamily 2209.00665}}].

\bibitem{Acanfora:2023gzr}
F.~Acanfora, R.~Franceschini, A.~Mastroddi and D.~Redigolo, \emph{{Fusing
  photons into nothing, a new search for invisible ALPs and Dark Matter at
  Belle II}},  \href{https://arxiv.org/abs/2307.06369}{{\ttfamily 2307.06369}}.

\bibitem{Belle-II:2023esi}
{\scshape Belle-II} collaboration, \emph{{Evidence for $B^+ \to K^+ \nu\bar\nu$
  decays}}, \href{https://doi.org/10.1103/PhysRevD.109.112006}{\emph{Phys. Rev.
  D} {\bfseries 109} (2024) 112006}
  [\href{https://arxiv.org/abs/2311.14647}{{\ttfamily 2311.14647}}].

\bibitem{Colangelo:1996ay}
P.~Colangelo, F.~De~Fazio, P.~Santorelli and E.~Scrimieri, \emph{{Rare $B \to
  K^{(*)}$ neutrino anti-neutrino decays at $B$ factories}},
  \href{https://doi.org/10.1016/S0370-2693(97)00130-5}{\emph{Phys. Lett. B}
  {\bfseries 395} (1997) 339}
  [\href{https://arxiv.org/abs/hep-ph/9610297}{{\ttfamily hep-ph/9610297}}].

\bibitem{Buchalla:2000sk}
G.~Buchalla, G.~Hiller and G.~Isidori, \emph{{Phenomenology of nonstandard $Z$
  couplings in exclusive semileptonic $b \to s$ transitions}},
  \href{https://doi.org/10.1103/PhysRevD.63.014015}{\emph{Phys. Rev. D}
  {\bfseries 63} (2000) 014015}
  [\href{https://arxiv.org/abs/hep-ph/0006136}{{\ttfamily hep-ph/0006136}}].

\bibitem{Altmannshofer:2009ma}
W.~Altmannshofer, A.J.~Buras, D.M.~Straub and M.~Wick, \emph{{New strategies
  for New Physics search in $B \to K^{*} \nu \bar{\nu}$, $B \to K \nu
  \bar{\nu}$ and $B \to X_{s} \nu \bar{\nu}$ decays}},
  \href{https://doi.org/10.1088/1126-6708/2009/04/022}{\emph{JHEP} {\bfseries
  04} (2009) 022} [\href{https://arxiv.org/abs/0902.0160}{{\ttfamily
  0902.0160}}].

\bibitem{Buras:2014fpa}
A.J.~Buras, J.~Girrbach-Noe, C.~Niehoff and D.M.~Straub, \emph{{$ B\to
  {K}^{\left(\ast \right)}\nu \overline{\nu} $ decays in the Standard Model and
  beyond}}, \href{https://doi.org/10.1007/JHEP02(2015)184}{\emph{JHEP}
  {\bfseries 02} (2015) 184} [\href{https://arxiv.org/abs/1409.4557}{{\ttfamily
  1409.4557}}].

\bibitem{Browder:2021hbl}
T.E.~Browder, N.G.~Deshpande, R.~Mandal and R.~Sinha, \emph{{Impact of
  B\textrightarrow{}K\ensuremath{\nu}\ensuremath{\nu}\textasciimacron{}
  measurements on beyond the Standard Model theories}},
  \href{https://doi.org/10.1103/PhysRevD.104.053007}{\emph{Phys. Rev. D}
  {\bfseries 104} (2021) 053007}
  [\href{https://arxiv.org/abs/2107.01080}{{\ttfamily 2107.01080}}].

\bibitem{Bause:2021cna}
R.~Bause, H.~Gisbert, M.~Golz and G.~Hiller, \emph{{Interplay of dineutrino
  modes with semileptonic rare B-decays}},
  \href{https://doi.org/10.1007/JHEP12(2021)061}{\emph{JHEP} {\bfseries 12}
  (2021) 061} [\href{https://arxiv.org/abs/2109.01675}{{\ttfamily
  2109.01675}}].

\bibitem{Felkl:2021uxi}
T.~Felkl, S.L.~Li and M.A.~Schmidt, \emph{{A tale of invisibility: constraints
  on new physics in b \textrightarrow{} s\ensuremath{\nu}\ensuremath{\nu}}},
  \href{https://doi.org/10.1007/JHEP12(2021)118}{\emph{JHEP} {\bfseries 12}
  (2021) 118} [\href{https://arxiv.org/abs/2111.04327}{{\ttfamily
  2111.04327}}].

\bibitem{Becirevic:2023aov}
D.~Be\v{c}irevi\'c, G.~Piazza and O.~Sumensari, \emph{{Revisiting $B\rightarrow
  K^{(*)} \nu {\bar{\nu }}$ decays in the Standard Model and beyond}},
  \href{https://doi.org/10.1140/epjc/s10052-023-11388-z}{\emph{Eur. Phys. J. C}
  {\bfseries 83} (2023) 252}
  [\href{https://arxiv.org/abs/2301.06990}{{\ttfamily 2301.06990}}].

\bibitem{Athron:2023hmz}
P.~Athron, R.~Martinez and C.~Sierra, \emph{{B meson anomalies and large $
  {B}^{+}\to {K}^{+}\nu \overline{\nu} $ in non-universal $U(1)^{'}$ models}},
  \href{https://doi.org/10.1007/JHEP02(2024)121}{\emph{JHEP} {\bfseries 02}
  (2024) 121} [\href{https://arxiv.org/abs/2308.13426}{{\ttfamily
  2308.13426}}].

\bibitem{Bause:2023mfe}
R.~Bause, H.~Gisbert and G.~Hiller, \emph{{Implications of an enhanced
  B\textrightarrow{}K\ensuremath{\nu}\ensuremath{\nu}\textasciimacron{}
  branching ratio}},
  \href{https://doi.org/10.1103/PhysRevD.109.015006}{\emph{Phys. Rev. D}
  {\bfseries 109} (2024) 015006}
  [\href{https://arxiv.org/abs/2309.00075}{{\ttfamily 2309.00075}}].

\bibitem{Allwicher:2023xba}
L.~Allwicher, D.~Becirevic, G.~Piazza, S.~Rosauro-Alcaraz and O.~Sumensari,
  \emph{{Understanding the first measurement of
  B(B\textrightarrow{}K\ensuremath{\nu}\ensuremath{\nu}\textasciimacron{})}},
  \href{https://doi.org/10.1016/j.physletb.2023.138411}{\emph{Phys. Lett. B}
  {\bfseries 848} (2024) 138411}
  [\href{https://arxiv.org/abs/2309.02246}{{\ttfamily 2309.02246}}].

\bibitem{Felkl:2023ayn}
T.~Felkl, A.~Giri, R.~Mohanta and M.A.~Schmidt, \emph{{When energy goes
  missing: new physics in $b\rightarrow s \nu \nu $ with sterile neutrinos}},
  \href{https://doi.org/10.1140/epjc/s10052-023-12326-9}{\emph{Eur. Phys. J. C}
  {\bfseries 83} (2023) 1135}
  [\href{https://arxiv.org/abs/2309.02940}{{\ttfamily 2309.02940}}].

\bibitem{Dreiner:2023cms}
H.K.~Dreiner, J.Y.~G\"unther and Z.S.~Wang, \emph{{The Decay $B\to
  K\nu\bar{\nu}$ at Belle II and a Massless Bino in R-parity-violating
  Supersymmetry}},  \href{https://arxiv.org/abs/2309.03727}{{\ttfamily
  2309.03727}}.

\bibitem{Abdughani:2023dlr}
M.~Abdughani and Y.~Reyimuaji, \emph{{Constraining light dark matter and
  mediator with $B^+ \to K^+ \nu\bar\nu$ data}},
  \href{https://doi.org/10.1103/PhysRevD.110.055013}{\emph{Phys. Rev. D}
  {\bfseries 110} (2024) 055013}
  [\href{https://arxiv.org/abs/2309.03706}{{\ttfamily 2309.03706}}].

\bibitem{He:2023bnk}
X.-G.~He, X.-D.~Ma and G.~Valencia, \emph{{Revisiting models that enhance $B^+
  \to K^+ \nu\bar\nu$ in light of the new Belle II measurement}},
  \href{https://doi.org/10.1103/PhysRevD.109.075019}{\emph{Phys. Rev. D}
  {\bfseries 109} (2024) 075019}
  [\href{https://arxiv.org/abs/2309.12741}{{\ttfamily 2309.12741}}].

\bibitem{Berezhnoy:2023rxx}
A.~Berezhnoy and D.~Melikhov, \emph{{$B\to K^* M_X$ vs $B\to K M_X$ as a probe
  of a scalar-mediator dark matter scenario}},
  \href{https://doi.org/10.1209/0295-5075/ad1d03}{\emph{EPL} {\bfseries 145}
  (2024) 14001} [\href{https://arxiv.org/abs/2309.17191}{{\ttfamily
  2309.17191}}].

\bibitem{Datta:2023iln}
A.~Datta, D.~Marfatia and L.~Mukherjee,
  \emph{{B\textrightarrow{}K\ensuremath{\nu}\ensuremath{\nu}\textasciimacron{},
  MiniBooNE and muon g-2 anomalies from a dark sector}},
  \href{https://doi.org/10.1103/PhysRevD.109.L031701}{\emph{Phys. Rev. D}
  {\bfseries 109} (2024) L031701}
  [\href{https://arxiv.org/abs/2310.15136}{{\ttfamily 2310.15136}}].

\bibitem{Altmannshofer:2023hkn}
W.~Altmannshofer, A.~Crivellin, H.~Haigh, G.~Inguglia and J.~Martin~Camalich,
  \emph{{Light new physics in
  B\textrightarrow{}K(*)\ensuremath{\nu}\ensuremath{\nu}\textasciimacron{}?}},
  \href{https://doi.org/10.1103/PhysRevD.109.075008}{\emph{Phys. Rev. D}
  {\bfseries 109} (2024) 075008}
  [\href{https://arxiv.org/abs/2311.14629}{{\ttfamily 2311.14629}}].

\bibitem{McKeen:2023uzo}
D.~McKeen, J.N.~Ng and D.~Tuckler, \emph{{Higgs portal interpretation of the
  Belle II B+\textrightarrow{}K+\ensuremath{\nu}\ensuremath{\nu} measurement}},
  \href{https://doi.org/10.1103/PhysRevD.109.075006}{\emph{Phys. Rev. D}
  {\bfseries 109} (2024) 075006}
  [\href{https://arxiv.org/abs/2312.00982}{{\ttfamily 2312.00982}}].

\bibitem{Fridell:2023ssf}
K.~Fridell, M.~Ghosh, T.~Okui and K.~Tobioka, \emph{{Decoding the
  B\textrightarrow{}K\ensuremath{\nu}\ensuremath{\nu} excess at Belle II:
  Kinematics, operators, and masses}},
  \href{https://doi.org/10.1103/PhysRevD.109.115006}{\emph{Phys. Rev. D}
  {\bfseries 109} (2024) 115006}
  [\href{https://arxiv.org/abs/2312.12507}{{\ttfamily 2312.12507}}].

\bibitem{Ho:2024cwk}
S.-Y.~Ho, J.~Kim and P.~Ko, \emph{{Recent $B^+ \!\to K^+\nu\bar{\nu}$ Excess
  and Muon $g-2$ Illuminating Light Dark Sector with Higgs Portal}},
  \href{https://arxiv.org/abs/2401.10112}{{\ttfamily 2401.10112}}.

\bibitem{Gabrielli:2024wys}
E.~Gabrielli, L.~Marzola, K.~M\"u\"ursepp and M.~Raidal, \emph{{Explaining the
  $B^+\rightarrow K^+ \nu \bar{\nu }$ excess via a massless dark photon}},
  \href{https://doi.org/10.1140/epjc/s10052-024-12818-2}{\emph{Eur. Phys. J. C}
  {\bfseries 84} (2024) 460}
  [\href{https://arxiv.org/abs/2402.05901}{{\ttfamily 2402.05901}}].

\bibitem{Li:2024thq}
T.~Li, Z.~Qian, M.A.~Schmidt and M.~Yuan, \emph{{The quark flavor-violating
  ALPs in light of B mesons and hadron colliders}},
  \href{https://doi.org/10.1007/JHEP05(2024)232}{\emph{JHEP} {\bfseries 05}
  (2024) 232} [\href{https://arxiv.org/abs/2402.14232}{{\ttfamily
  2402.14232}}].

\bibitem{Hou:2024vyw}
B.-F.~Hou, X.-Q.~Li, M.~Shen, Y.-D.~Yang and X.-B.~Yuan, \emph{{Deciphering the
  Belle II data on $ B\to K\nu \overline{\nu} $ decay in the (dark) SMEFT with
  minimal flavour violation}},
  \href{https://doi.org/10.1007/JHEP06(2024)172}{\emph{JHEP} {\bfseries 06}
  (2024) 172} [\href{https://arxiv.org/abs/2402.19208}{{\ttfamily
  2402.19208}}].

\bibitem{He:2024iju}
X.-G.~He, X.-D.~Ma, M.A.~Schmidt, G.~Valencia and R.R.~Volkas, \emph{{Scalar
  dark matter explanation of the excess in the Belle II
  B$^{+}$\textrightarrow{} K$^{+}$+ invisible measurement}},
  \href{https://doi.org/10.1007/JHEP07(2024)168}{\emph{JHEP} {\bfseries 07}
  (2024) 168} [\href{https://arxiv.org/abs/2403.12485}{{\ttfamily
  2403.12485}}].

\bibitem{Bolton:2024egx}
P.D.~Bolton, S.~Fajfer, J.F.~Kamenik and M.~Novoa-Brunet, \emph{{Signatures of
  light new particles in B\textrightarrow{}K(*)Emiss}},
  \href{https://doi.org/10.1103/PhysRevD.110.055001}{\emph{Phys. Rev. D}
  {\bfseries 110} (2024) 055001}
  [\href{https://arxiv.org/abs/2403.13887}{{\ttfamily 2403.13887}}].

\bibitem{Marzocca:2024hua}
D.~Marzocca, M.~Nardecchia, A.~Stanzione and C.~Toni, \emph{{Implications of $B
  \to K \nu \bar{\nu}$ under Rank-One Flavor Violation hypothesis}},
  \href{https://arxiv.org/abs/2404.06533}{{\ttfamily 2404.06533}}.

\bibitem{Aghaie:2024jkj}
M.~Aghaie, G.~Armando, A.~Conaci, A.~Dondarini, P.~Matak, P.~Panci et~al.,
  \emph{{Axion dark matter from heavy quarks}},
  \href{https://doi.org/10.1016/j.physletb.2024.138923}{\emph{Phys. Lett. B}
  {\bfseries 856} (2024) 138923}
  [\href{https://arxiv.org/abs/2404.12199}{{\ttfamily 2404.12199}}].

\bibitem{Rosauro-Alcaraz:2024mvx}
S.~Rosauro-Alcaraz and L.P.S.~Leal, \emph{{Disentangling left and right-handed
  neutrino effects in $B\rightarrow K^{(*)}\nu \nu $}},
  \href{https://doi.org/10.1140/epjc/s10052-024-13104-x}{\emph{Eur. Phys. J. C}
  {\bfseries 84} (2024) 795}
  [\href{https://arxiv.org/abs/2404.17440}{{\ttfamily 2404.17440}}].

\bibitem{Eguren:2024oov}
J.F.~Eguren, S.~Klingel, E.~Stamou, M.~Tabet and R.~Ziegler, \emph{{Flavor
  phenomenology of light dark vectors}},
  \href{https://doi.org/10.1007/JHEP08(2024)111}{\emph{JHEP} {\bfseries 08}
  (2024) 111} [\href{https://arxiv.org/abs/2405.00108}{{\ttfamily
  2405.00108}}].

\bibitem{Buras:2024ewl}
A.J.~Buras, J.~Harz and M.A.~Mojahed, \emph{{Disentangling new physics in $
  K\to \pi \nu \overline{\nu} $ and $ B\to K\left({K}^{\ast}\right)\nu
  \overline{\nu} $ observables}},
  \href{https://doi.org/10.1007/JHEP10(2024)087}{\emph{JHEP} {\bfseries 10}
  (2024) 087} [\href{https://arxiv.org/abs/2405.06742}{{\ttfamily
  2405.06742}}].

\bibitem{Hati:2024ppg}
C.~Hati, J.~Leite, N.~Nath and J.W.F.~Valle, \emph{{The QCD axion,
  colour-mediated neutrino masses, and $B^+\to K^+ + E_{\text{miss}}$
  anomaly}},  \href{https://arxiv.org/abs/2408.00060}{{\ttfamily 2408.00060}}.

\bibitem{Wang:2024prt}
Z.S.~Wang, Y.~Zhang and W.~Liu, \emph{{Long-lived sterile neutrinos from an
  axionlike particle at Belle II}},
  \href{https://arxiv.org/abs/2410.00491}{{\ttfamily 2410.00491}}.

\bibitem{BaBar:2021ich}
{\scshape BaBar} collaboration, \emph{{Search for an Axionlike Particle in $B$
  Meson Decays}},
  \href{https://doi.org/10.1103/PhysRevLett.128.131802}{\emph{Phys. Rev. Lett.}
  {\bfseries 128} (2022) 131802}
  [\href{https://arxiv.org/abs/2111.01800}{{\ttfamily 2111.01800}}].

\bibitem{Bhattacharya:2021shk}
B.~Bhattacharya, A.~Datta, D.~Marfatia, S.~Nandi and J.~Waite,
  \emph{{Axion-like particles resolve the $B \to \pi K$ and $g-2$ anomalies}},
  \href{https://doi.org/10.1103/PhysRevD.104.L051701}{\emph{Phys. Rev. D}
  {\bfseries 104} (2021) L051701}
  [\href{https://arxiv.org/abs/2104.03947}{{\ttfamily 2104.03947}}].

\bibitem{Gronau:1994rj}
M.~Gronau, O.F.~Hernandez, D.~London and J.L.~Rosner, \emph{{Decays of $B$
  mesons to two light pseudoscalars}},
  \href{https://doi.org/10.1103/PhysRevD.50.4529}{\emph{Phys. Rev. D}
  {\bfseries 50} (1994) 4529}
  [\href{https://arxiv.org/abs/hep-ph/9404283}{{\ttfamily hep-ph/9404283}}].

\bibitem{Gronau:1995hn}
M.~Gronau, O.F.~Hernandez, D.~London and J.L.~Rosner, \emph{{Electroweak
  penguins and two-body $B$ decays}},
  \href{https://doi.org/10.1103/PhysRevD.52.6374}{\emph{Phys. Rev. D}
  {\bfseries 52} (1995) 6374}
  [\href{https://arxiv.org/abs/hep-ph/9504327}{{\ttfamily hep-ph/9504327}}].

\bibitem{Fleischer:1997um}
R.~Fleischer and T.~Mannel, \emph{{Constraining the CKM angle $\gamma$ and
  penguin contributions through combined $B\to\pi K$ branching ratios}},
  \href{https://doi.org/10.1103/PhysRevD.57.2752}{\emph{Phys. Rev. D}
  {\bfseries 57} (1998) 2752}
  [\href{https://arxiv.org/abs/hep-ph/9704423}{{\ttfamily hep-ph/9704423}}].

\bibitem{Gronau:1997an}
M.~Gronau and J.L.~Rosner, \emph{{Weak phase $\gamma$ from ratio of $B\to K\pi$
  rates}}, \href{https://doi.org/10.1103/PhysRevD.57.6843}{\emph{Phys. Rev. D}
  {\bfseries 57} (1998) 6843}
  [\href{https://arxiv.org/abs/hep-ph/9711246}{{\ttfamily hep-ph/9711246}}].

\bibitem{Neubert:1998pt}
M.~Neubert and J.L.~Rosner, \emph{{New bound on $\gamma$ from $B^{\pm} \to \pi
  K$ decays}}, \href{https://doi.org/10.1016/S0370-2693(98)01194-0}{\emph{Phys.
  Lett. B} {\bfseries 441} (1998) 403}
  [\href{https://arxiv.org/abs/hep-ph/9808493}{{\ttfamily hep-ph/9808493}}].

\bibitem{Neubert:1998jq}
M.~Neubert and J.L.~Rosner, \emph{{Determination of the weak phase $\gamma$
  from rate measurements in $B^{\pm} \to\pi K$, $\pi \pi$ decays}},
  \href{https://doi.org/10.1103/PhysRevLett.81.5076}{\emph{Phys. Rev. Lett.}
  {\bfseries 81} (1998) 5076}
  [\href{https://arxiv.org/abs/hep-ph/9809311}{{\ttfamily hep-ph/9809311}}].

\bibitem{Gronau:2006xu}
M.~Gronau and J.L.~Rosner, \emph{{Rate and CP-asymmetry sum rules in $B\to
  K\pi$}}, \href{https://doi.org/10.1103/PhysRevD.74.057503}{\emph{Phys. Rev.
  D} {\bfseries 74} (2006) 057503}
  [\href{https://arxiv.org/abs/hep-ph/0608040}{{\ttfamily hep-ph/0608040}}].

\bibitem{Buchalla:1995vs}
G.~Buchalla, A.J.~Buras and M.E.~Lautenbacher, \emph{{Weak decays beyond
  leading logarithms}},
  \href{https://doi.org/10.1103/RevModPhys.68.1125}{\emph{Rev. Mod. Phys.}
  {\bfseries 68} (1996) 1125}
  [\href{https://arxiv.org/abs/hep-ph/9512380}{{\ttfamily hep-ph/9512380}}].

\bibitem{Gronau:1998fn}
M.~Gronau, D.~Pirjol and T.-M.~Yan, \emph{{Model independent electroweak
  penguins in $B$ decays to two pseudoscalars}},
  \href{https://doi.org/10.1103/PhysRevD.60.034021}{\emph{Phys. Rev. D}
  {\bfseries 60} (1999) 034021}
  [\href{https://arxiv.org/abs/hep-ph/9810482}{{\ttfamily hep-ph/9810482}}].

\bibitem{Chau:1990ay}
L.-L.~Chau, H.-Y.~Cheng, W.K.~Sze, H.~Yao and B.~Tseng, \emph{{Charmless
  nonleptonic rare decays of $B$ mesons}},
  \href{https://doi.org/10.1103/PhysRevD.43.2176}{\emph{Phys. Rev. D}
  {\bfseries 43} (1991) 2176}.

\bibitem{Gronau:2005kz}
M.~Gronau, \emph{{A Precise sum rule among four $B\to K\pi$ CP asymmetries}},
  \href{https://doi.org/10.1016/j.physletb.2005.09.014}{\emph{Phys. Lett. B}
  {\bfseries 627} (2005) 82}
  [\href{https://arxiv.org/abs/hep-ph/0508047}{{\ttfamily hep-ph/0508047}}].

\bibitem{Gronau:1995hm}
M.~Gronau, O.F.~Hernandez, D.~London and J.L.~Rosner, \emph{{Broken SU(3)
  symmetry in two-body $B$ decays}},
  \href{https://doi.org/10.1103/PhysRevD.52.6356}{\emph{Phys. Rev. D}
  {\bfseries 52} (1995) 6356}
  [\href{https://arxiv.org/abs/hep-ph/9504326}{{\ttfamily hep-ph/9504326}}].

\bibitem{Keum:2000ms}
Y.-Y.~Keum and H.-n.~Li, \emph{{Nonleptonic charmless B decays: Factorization
  versus perturbative QCD}},
  \href{https://doi.org/10.1103/PhysRevD.63.074006}{\emph{Phys. Rev. D}
  {\bfseries 63} (2001) 074006}
  [\href{https://arxiv.org/abs/hep-ph/0006001}{{\ttfamily hep-ph/0006001}}].

\bibitem{Beneke:2000ry}
M.~Beneke, G.~Buchalla, M.~Neubert and C.T.~Sachrajda, \emph{{QCD factorization
  for exclusive, nonleptonic $B$ meson decays: General arguments and the case
  of heavy light final states}},
  \href{https://doi.org/10.1016/S0550-3213(00)00559-9}{\emph{Nucl. Phys. B}
  {\bfseries 591} (2000) 313}
  [\href{https://arxiv.org/abs/hep-ph/0006124}{{\ttfamily hep-ph/0006124}}].

\bibitem{Beneke:2001ev}
M.~Beneke, G.~Buchalla, M.~Neubert and C.T.~Sachrajda, \emph{{QCD factorization
  in $B \to \pi K$, $\pi \pi$ decays and extraction of Wolfenstein
  parameters}},
  \href{https://doi.org/10.1016/S0550-3213(01)00251-6}{\emph{Nucl. Phys. B}
  {\bfseries 606} (2001) 245}
  [\href{https://arxiv.org/abs/hep-ph/0104110}{{\ttfamily hep-ph/0104110}}].

\bibitem{Li:2009wba}
H.-n.~Li and S.~Mishima, \emph{{Possible resolution of the $B\to\pi \pi$, $\pi
  K$ puzzles}}, \href{https://doi.org/10.1103/PhysRevD.83.034023}{\emph{Phys.
  Rev. D} {\bfseries 83} (2011) 034023}
  [\href{https://arxiv.org/abs/0901.1272}{{\ttfamily 0901.1272}}].

\bibitem{Bauer:2005kd}
C.W.~Bauer, I.Z.~Rothstein and I.W.~Stewart, \emph{{SCET analysis of $B\to
  K\pi$, $B\to K \bar{K}$, and $B\to\pi \pi$ decays}},
  \href{https://doi.org/10.1103/PhysRevD.74.034010}{\emph{Phys. Rev. D}
  {\bfseries 74} (2006) 034010}
  [\href{https://arxiv.org/abs/hep-ph/0510241}{{\ttfamily hep-ph/0510241}}].

\bibitem{Huitu:2009st}
K.~Huitu and S.~Khalil, \emph{{New Physics contribution to $B\to K\pi$ decays
  in SCET}}, \href{https://doi.org/10.1103/PhysRevD.81.095008}{\emph{Phys. Rev.
  D} {\bfseries 81} (2010) 095008}
  [\href{https://arxiv.org/abs/0911.1868}{{\ttfamily 0911.1868}}].

\bibitem{Gronau:1998ep}
M.~Gronau and J.L.~Rosner, \emph{{Combining CP asymmetries in $B \to K\pi$
  decays}}, \href{https://doi.org/10.1103/PhysRevD.59.113002}{\emph{Phys. Rev.
  D} {\bfseries 59} (1999) 113002}
  [\href{https://arxiv.org/abs/hep-ph/9809384}{{\ttfamily hep-ph/9809384}}].

\bibitem{Duh:2012ie}
{\scshape Belle} collaboration, \emph{{Measurements of branching fractions and
  direct CP asymmetries for $B\to K \pi$, $B\to \pi \pi$ and $B\to K K$
  decays}}, \href{https://doi.org/10.1103/PhysRevD.87.031103}{\emph{Phys. Rev.
  D} {\bfseries 87} (2013) 031103}
  [\href{https://arxiv.org/abs/1210.1348}{{\ttfamily 1210.1348}}].

\bibitem{Aaij:2020wnj}
{\scshape LHCb} collaboration, \emph{{Measurement of CP Violation in the Decay
  $B^{+} \rightarrow K^{+} \pi^{0}$}},
  \href{https://doi.org/10.1103/PhysRevLett.126.091802}{\emph{Phys. Rev. Lett.}
  {\bfseries 126} (2021) 091802}
  [\href{https://arxiv.org/abs/2012.12789}{{\ttfamily 2012.12789}}].

\bibitem{Belle-II:2023ksq}
{\scshape Belle-II} collaboration, \emph{{Measurement of branching fractions
  and direct CP asymmetries for B\textrightarrow{}K\ensuremath{\pi} and
  B\textrightarrow{}\ensuremath{\pi}\ensuremath{\pi} decays at Belle II}},
  \href{https://doi.org/10.1103/PhysRevD.109.012001}{\emph{Phys. Rev. D}
  {\bfseries 109} (2024) 012001}
  [\href{https://arxiv.org/abs/2310.06381}{{\ttfamily 2310.06381}}].

\bibitem{Bobeth:2014rra}
C.~Bobeth, M.~Gorbahn and S.~Vickers, \emph{{Weak annihilation and new physics
  in charmless $B \to M M$ decays}},
  \href{https://doi.org/10.1140/epjc/s10052-015-3535-1}{\emph{Eur. Phys. J. C}
  {\bfseries 75} (2015) 340} [\href{https://arxiv.org/abs/1409.3252}{{\ttfamily
  1409.3252}}].

\bibitem{Huber:2021cgk}
T.~Huber and G.~Tetlalmatzi-Xolocotzi, \emph{{Estimating QCD-factorization
  amplitudes through SU(3) symmetry in $B\rightarrow P P$ decays}},
  \href{https://doi.org/10.1140/epjc/s10052-022-10068-8}{\emph{Eur. Phys. J. C}
  {\bfseries 82} (2022) 210}
  [\href{https://arxiv.org/abs/2111.06418}{{\ttfamily 2111.06418}}].

\bibitem{Baek:2004rp}
S.~Baek, P.~Hamel, D.~London, A.~Datta and D.A.~Suprun, \emph{{The $B \to \pi
  K$ puzzle and new physics}},
  \href{https://doi.org/10.1103/PhysRevD.71.057502}{\emph{Phys. Rev. D}
  {\bfseries 71} (2005) 057502}
  [\href{https://arxiv.org/abs/hep-ph/0412086}{{\ttfamily hep-ph/0412086}}].

\bibitem{Datta:2004re}
A.~Datta and D.~London, \emph{{Measuring new physics parameters in $B$ penguin
  decays}}, \href{https://doi.org/10.1016/j.physletb.2004.06.069}{\emph{Phys.
  Lett. B} {\bfseries 595} (2004) 453}
  [\href{https://arxiv.org/abs/hep-ph/0404130}{{\ttfamily hep-ph/0404130}}].

\bibitem{Neubert:1997wb}
M.~Neubert, \emph{{Rescattering effects, isospin relations and electroweak
  penguins in $B \to \pi K$ decays}},
  \href{https://doi.org/10.1016/S0370-2693(98)00175-0}{\emph{Phys. Lett. B}
  {\bfseries 424} (1998) 152}
  [\href{https://arxiv.org/abs/hep-ph/9712224}{{\ttfamily hep-ph/9712224}}].

\bibitem{Atwood:1997iw}
D.~Atwood and A.~Soni, \emph{{The Possibility of large direct CP violation in
  $B \to K\pi$ - like modes due to long distance rescattering effects and
  implications for the angle $\gamma$}},
  \href{https://doi.org/10.1103/PhysRevD.58.036005}{\emph{Phys. Rev. D}
  {\bfseries 58} (1998) 036005}
  [\href{https://arxiv.org/abs/hep-ph/9712287}{{\ttfamily hep-ph/9712287}}].

\bibitem{Buras:1997cv}
A.J.~Buras, R.~Fleischer and T.~Mannel, \emph{{Penguin topologies, rescattering
  effects and penguin hunting with $B_{u,d} \to K \bar{K}$ and $B^{\pm} \to
  \pi^{\pm} K$}},
  \href{https://doi.org/10.1016/S0550-3213(98)00506-9}{\emph{Nucl. Phys. B}
  {\bfseries 533} (1998) 3}
  [\href{https://arxiv.org/abs/hep-ph/9711262}{{\ttfamily hep-ph/9711262}}].

\bibitem{Buras:2000gc}
A.J.~Buras and R.~Fleischer, \emph{{Constraints on the CKM angle $\gamma$ and
  strong phases from $B \to \pi K$ decays}},
  \href{https://doi.org/10.1007/s100520050006}{\emph{Eur. Phys. J. C}
  {\bfseries 16} (2000) 97}
  [\href{https://arxiv.org/abs/hep-ph/0003323}{{\ttfamily hep-ph/0003323}}].

\bibitem{Falk:1998wc}
A.F.~Falk, A.L.~Kagan, Y.~Nir and A.A.~Petrov, \emph{{Final state interactions
  and new physics in $B \to \pi K$ decays}},
  \href{https://doi.org/10.1103/PhysRevD.57.4290}{\emph{Phys. Rev. D}
  {\bfseries 57} (1998) 4290}
  [\href{https://arxiv.org/abs/hep-ph/9712225}{{\ttfamily hep-ph/9712225}}].

\bibitem{Cheng:2004ru}
H.-Y.~Cheng, C.-K.~Chua and A.~Soni, \emph{{Final state interactions in
  hadronic B decays}},
  \href{https://doi.org/10.1103/PhysRevD.71.014030}{\emph{Phys. Rev. D}
  {\bfseries 71} (2005) 014030}
  [\href{https://arxiv.org/abs/hep-ph/0409317}{{\ttfamily hep-ph/0409317}}].

\bibitem{Buras:1998ra}
A.J.~Buras and L.~Silvestrini, \emph{{Nonleptonic two-body $B$ decays beyond
  factorization}},
  \href{https://doi.org/10.1016/S0550-3213(99)00712-9}{\emph{Nucl. Phys. B}
  {\bfseries 569} (2000) 3}
  [\href{https://arxiv.org/abs/hep-ph/9812392}{{\ttfamily hep-ph/9812392}}].

\bibitem{Bauer:2004ck}
C.W.~Bauer and D.~Pirjol, \emph{{Graphical amplitudes from SCET}},
  \href{https://doi.org/10.1016/j.physletb.2004.10.047}{\emph{Phys. Lett. B}
  {\bfseries 604} (2004) 183}
  [\href{https://arxiv.org/abs/hep-ph/0408161}{{\ttfamily hep-ph/0408161}}].

\bibitem{Bauer:2004tj}
C.W.~Bauer, D.~Pirjol, I.Z.~Rothstein and I.W.~Stewart, \emph{{$B \to M_1 M_2$:
  Factorization, charming penguins, strong phases, and polarization}},
  \href{https://doi.org/10.1103/PhysRevD.70.054015}{\emph{Phys. Rev. D}
  {\bfseries 70} (2004) 054015}
  [\href{https://arxiv.org/abs/hep-ph/0401188}{{\ttfamily hep-ph/0401188}}].

\bibitem{Kamal:1999rn}
A.N.~Kamal, \emph{{Long range final state interactions and direct CP asymmetry
  in B+ ---\ensuremath{>} pi+ K0 decay}},
  \href{https://doi.org/10.1103/PhysRevD.60.094018}{\emph{Phys. Rev. D}
  {\bfseries 60} (1999) 094018}
  [\href{https://arxiv.org/abs/hep-ph/9901342}{{\ttfamily hep-ph/9901342}}].

\bibitem{Cheng:2005bg}
H.-Y.~Cheng, C.-K.~Chua and A.~Soni, \emph{{Effects of final-state interactions
  on mixing-induced CP violation in penguin-dominated B decays}},
  \href{https://doi.org/10.1103/PhysRevD.72.014006}{\emph{Phys. Rev. D}
  {\bfseries 72} (2005) 014006}
  [\href{https://arxiv.org/abs/hep-ph/0502235}{{\ttfamily hep-ph/0502235}}].

\bibitem{Ciuchini:2001gv}
M.~Ciuchini, E.~Franco, G.~Martinelli, M.~Pierini and L.~Silvestrini,
  \emph{{Charming penguins strike back}},
  \href{https://doi.org/10.1016/S0370-2693(01)00700-6}{\emph{Phys. Lett. B}
  {\bfseries 515} (2001) 33}
  [\href{https://arxiv.org/abs/hep-ph/0104126}{{\ttfamily hep-ph/0104126}}].

\bibitem{Ciuchini:2008eh}
M.~Ciuchini, E.~Franco, G.~Martinelli, M.~Pierini and L.~Silvestrini,
  \emph{{Searching For New Physics With $B \to K\pi$ Decays}},
  \href{https://doi.org/10.1016/j.physletb.2009.03.011}{\emph{Phys. Lett. B}
  {\bfseries 674} (2009) 197}
  [\href{https://arxiv.org/abs/0811.0341}{{\ttfamily 0811.0341}}].

\bibitem{Beaudry:2017gtw}
N.B.~Beaudry, A.~Datta, D.~London, A.~Rashed and J.-S.~Roux, \emph{{The $B \to
  \pi K$ puzzle revisited}},
  \href{https://doi.org/10.1007/JHEP01(2018)074}{\emph{JHEP} {\bfseries 01}
  (2018) 074} [\href{https://arxiv.org/abs/1709.07142}{{\ttfamily
  1709.07142}}].

\bibitem{Fleischer:2017vrb}
R.~Fleischer, R.~Jaarsma and K.K.~Vos, \emph{{Towards new frontiers with
  $B\to\pi K$ decays}},
  \href{https://doi.org/10.1016/j.physletb.2018.09.015}{\emph{Phys. Lett. B}
  {\bfseries 785} (2018) 525}
  [\href{https://arxiv.org/abs/1712.02323}{{\ttfamily 1712.02323}}].

\bibitem{Fleischer:2018bld}
R.~Fleischer, R.~Jaarsma, E.~Malami and K.K.~Vos, \emph{{Exploring
  $B\rightarrow \pi \pi , \pi K$ decays at the high-precision frontier}},
  \href{https://doi.org/10.1140/epjc/s10052-018-6397-5}{\emph{Eur. Phys. J. C}
  {\bfseries 78} (2018) 943}
  [\href{https://arxiv.org/abs/1806.08783}{{\ttfamily 1806.08783}}].

\bibitem{Kundu:2021emt}
A.~Kundu, S.K.~Patra and S.~Roy, \emph{{Complete analysis of all
  B\textrightarrow{}\ensuremath{\pi}K decays}},
  \href{https://doi.org/10.1103/PhysRevD.104.095025}{\emph{Phys. Rev. D}
  {\bfseries 104} (2021) 095025}
  [\href{https://arxiv.org/abs/2106.15633}{{\ttfamily 2106.15633}}].

\bibitem{Keum:2000wi}
Y.Y.~Keum, H.-N.~Li and A.I.~Sanda, \emph{{Penguin enhancement and $B \to K
  \pi$ decays in perturbative QCD}},
  \href{https://doi.org/10.1103/PhysRevD.63.054008}{\emph{Phys. Rev. D}
  {\bfseries 63} (2001) 054008}
  [\href{https://arxiv.org/abs/hep-ph/0004173}{{\ttfamily hep-ph/0004173}}].

\bibitem{Beneke:2003zv}
M.~Beneke and M.~Neubert, \emph{{QCD factorization for $B \to P P$ and $B \to P
  V$ decays}},
  \href{https://doi.org/10.1016/j.nuclphysb.2003.09.026}{\emph{Nucl. Phys. B}
  {\bfseries 675} (2003) 333}
  [\href{https://arxiv.org/abs/hep-ph/0308039}{{\ttfamily hep-ph/0308039}}].

\bibitem{Chang:2008tf}
Q.~Chang, X.-Q.~Li and Y.-D.~Yang, \emph{{Revisiting $B\to\pi K$, $\pi K^*$ and
  $\rho K$ Decays: Direct CP Violation and Implication for New Physics}},
  \href{https://doi.org/10.1088/1126-6708/2008/09/038}{\emph{JHEP} {\bfseries
  09} (2008) 038} [\href{https://arxiv.org/abs/0807.4295}{{\ttfamily
  0807.4295}}].

\bibitem{Cheng:2009eg}
H.-Y.~Cheng and C.-K.~Chua, \emph{{Resolving $B$-CP Puzzles in QCD
  Factorization}},
  \href{https://doi.org/10.1103/PhysRevD.80.074031}{\emph{Phys. Rev. D}
  {\bfseries 80} (2009) 074031}
  [\href{https://arxiv.org/abs/0908.3506}{{\ttfamily 0908.3506}}].

\bibitem{Li:2014haa}
H.-n.~Li and S.~Mishima, \emph{{Glauber gluons in spectator amplitudes for $B
  \to \pi M$ decays}},
  \href{https://doi.org/10.1103/PhysRevD.90.074018}{\emph{Phys. Rev. D}
  {\bfseries 90} (2014) 074018}
  [\href{https://arxiv.org/abs/1407.7647}{{\ttfamily 1407.7647}}].

\bibitem{Liu:2015upa}
X.~Liu, H.-n.~Li and Z.-J.~Xiao, \emph{{Resolving the $B\to K\pi$ puzzle by
  Glauber-gluon effects}},
  \href{https://doi.org/10.1103/PhysRevD.93.014024}{\emph{Phys. Rev. D}
  {\bfseries 93} (2016) 014024}
  [\href{https://arxiv.org/abs/1510.05910}{{\ttfamily 1510.05910}}].

\bibitem{Li:2005kt}
H.-n.~Li, S.~Mishima and A.I.~Sanda, \emph{{Resolution to the $B\to\pi K$
  puzzle}}, \href{https://doi.org/10.1103/PhysRevD.72.114005}{\emph{Phys. Rev.
  D} {\bfseries 72} (2005) 114005}
  [\href{https://arxiv.org/abs/hep-ph/0508041}{{\ttfamily hep-ph/0508041}}].

\bibitem{Kim:2007kx}
C.S.~Kim, S.~Oh and Y.W.~Yoon, \emph{{Analytic resolution of puzzle in $B \to
  K\pi$ decays}},
  \href{https://doi.org/10.1016/j.physletb.2008.06.015}{\emph{Phys. Lett. B}
  {\bfseries 665} (2008) 231}
  [\href{https://arxiv.org/abs/0707.2967}{{\ttfamily 0707.2967}}].

\bibitem{Buras:2002yj}
A.J.~Buras, F.~Parodi and A.~Stocchi, \emph{{The CKM matrix and the unitarity
  triangle: Another look}},
  \href{https://doi.org/10.1088/1126-6708/2003/01/029}{\emph{JHEP} {\bfseries
  01} (2003) 029} [\href{https://arxiv.org/abs/hep-ph/0207101}{{\ttfamily
  hep-ph/0207101}}].

\bibitem{Belle-II:2024eob}
{\scshape Belle-II, Belle} collaboration, \emph{{Determination of the CKM angle
  $\phi_{3}$ from a combination of Belle and Belle II results}},
  \href{https://arxiv.org/abs/2404.12817}{{\ttfamily 2404.12817}}.

\bibitem{Ball:2004ye}
P.~Ball and R.~Zwicky, \emph{{New results on $B \to \pi, K, \eta$ decay
  formfactors from light-cone sum rules}},
  \href{https://doi.org/10.1103/PhysRevD.71.014015}{\emph{Phys. Rev. D}
  {\bfseries 71} (2005) 014015}
  [\href{https://arxiv.org/abs/hep-ph/0406232}{{\ttfamily hep-ph/0406232}}].

\bibitem{Gubernari:2023puw}
N.~Gubernari, M.~Reboud, D.~van Dyk and J.~Virto, \emph{{Dispersive analysis of
  $B \to K^{(*)}$ and $B_{s} \to \phi$ form factors}},
  \href{https://doi.org/10.1007/JHEP12(2023)153}{\emph{JHEP} {\bfseries 12}
  (2023) 153} [\href{https://arxiv.org/abs/2305.06301}{{\ttfamily
  2305.06301}}].

\bibitem{Parrott:2022smq}
W.~Parrott, C.~Bouchard and C.~Davies, \emph{{The search for new physics in $B
  \to K \ell^+\ell^-$ and $B \to K \nu\bar{\nu}$ using precise lattice QCD form
  factors}}, \href{https://doi.org/10.22323/1.430.0421}{\emph{PoS} {\bfseries
  LATTICE2022} (2023) 421} [\href{https://arxiv.org/abs/2210.10898}{{\ttfamily
  2210.10898}}].

\bibitem{Bauer:2020jbp}
M.~Bauer, M.~Neubert, S.~Renner, M.~Schnubel and A.~Thamm, \emph{{The
  Low-Energy Effective Theory of Axions and ALPs}},
  \href{https://doi.org/10.1007/JHEP04(2021)063}{\emph{JHEP} {\bfseries 04}
  (2021) 063} [\href{https://arxiv.org/abs/2012.12272}{{\ttfamily
  2012.12272}}].

\bibitem{heidelbachmsc:22}
A.~Heidelbach, \emph{{{Sensitivity Study in the Search for $B^{\pm}\to
  K^{\pm}a$ (displaced $a\to\gamma\gamma$) decays at Belle II}}}, .

\bibitem{BaBar:2013npw}
{\scshape BaBar} collaboration, \emph{{Search for $B \to K^{(*)} \nu \overline
  \nu$ and invisible quarkonium decays}},
  \href{https://doi.org/10.1103/PhysRevD.87.112005}{\emph{Phys. Rev. D}
  {\bfseries 87} (2013) 112005}
  [\href{https://arxiv.org/abs/1303.7465}{{\ttfamily 1303.7465}}].

\bibitem{CHARM:1985nku}
{\scshape CHARM} collaboration, \emph{{A Search for Decays of Heavy Neutrinos
  in the Mass Range 0.5-{GeV} to 2.8-{GeV}}},
  \href{https://doi.org/10.1016/0370-2693(86)91601-1}{\emph{Phys. Lett. B}
  {\bfseries 166} (1986) 473}.

\bibitem{CHARM:1985anb}
{\scshape CHARM} collaboration, \emph{{Search for Axion Like Particle
  Production in 400-{GeV} Proton - Copper Interactions}},
  \href{https://doi.org/10.1016/0370-2693(85)90400-9}{\emph{Phys. Lett. B}
  {\bfseries 157} (1985) 458}.

\bibitem{Blumlein:1990ay}
J.~Blumlein et~al., \emph{{Limits on neutral light scalar and pseudoscalar
  particles in a proton beam dump experiment}},
  \href{https://doi.org/10.1007/BF01548556}{\emph{Z. Phys. C} {\bfseries 51}
  (1991) 341}.

\bibitem{Duffy:1988rw}
M.E.~Duffy et~al., \emph{{Neutrino Production by 400-{GeV}/c Protons in a
  Beam-dump Experiment}},
  \href{https://doi.org/10.1103/PhysRevD.38.2032}{\emph{Phys. Rev. D}
  {\bfseries 38} (1988) 2032}.

\bibitem{Ovchynnikov:2023cry}
M.~Ovchynnikov, J.-L.~Tastet, O.~Mikulenko and K.~Bondarenko,
  \emph{{Sensitivities to feebly interacting particles: Public and unified
  calculations}},
  \href{https://doi.org/10.1103/PhysRevD.108.075028}{\emph{Phys. Rev. D}
  {\bfseries 108} (2023) 075028}
  [\href{https://arxiv.org/abs/2305.13383}{{\ttfamily 2305.13383}}].

\bibitem{Abramowicz:2021zja}
H.~Abramowicz et~al., \emph{{Conceptual design report for the LUXE
  experiment}},
  \href{https://doi.org/10.1140/epjs/s11734-021-00249-z}{\emph{Eur. Phys. J.
  ST} {\bfseries 230} (2021) 2445}
  [\href{https://arxiv.org/abs/2102.02032}{{\ttfamily 2102.02032}}].

\bibitem{Bai:2021gbm}
Z.~Bai et~al., \emph{{New physics searches with an optical dump at LUXE}},
  \href{https://doi.org/10.1103/PhysRevD.106.115034}{\emph{Phys. Rev. D}
  {\bfseries 106} (2022) 115034}
  [\href{https://arxiv.org/abs/2107.13554}{{\ttfamily 2107.13554}}].

\bibitem{Alekhin:2015byh}
S.~Alekhin et~al., \emph{{A facility to Search for Hidden Particles at the CERN
  SPS: the SHiP physics case}},
  \href{https://doi.org/10.1088/0034-4885/79/12/124201}{\emph{Rept. Prog.
  Phys.} {\bfseries 79} (2016) 124201}
  [\href{https://arxiv.org/abs/1504.04855}{{\ttfamily 1504.04855}}].

\bibitem{Feng:2018pew}
J.L.~Feng, I.~Galon, F.~Kling and S.~Trojanowski, \emph{{Axionlike particles at
  FASER: The LHC as a photon beam dump}},
  \href{https://doi.org/10.1103/PhysRevD.98.055021}{\emph{Phys. Rev. D}
  {\bfseries 98} (2018) 055021}
  [\href{https://arxiv.org/abs/1806.02348}{{\ttfamily 1806.02348}}].

\bibitem{AxionLimits}
C.~O'Hare, ``cajohare/axionlimits: Axionlimits.''
  \url{https://cajohare.github.io/AxionLimits/}, July, 2020.
\newblock 10.5281/zenodo.3932430.

\bibitem{Lourenco:2006vw}
C.~Lourenco and H.K.~Wohri, \emph{{Heavy flavour hadro-production from
  fixed-target to collider energies}},
  \href{https://doi.org/10.1016/j.physrep.2006.05.005}{\emph{Phys. Rept.}
  {\bfseries 433} (2006) 127}
  [\href{https://arxiv.org/abs/hep-ph/0609101}{{\ttfamily hep-ph/0609101}}].

\bibitem{Clarke:2013aya}
J.D.~Clarke, R.~Foot and R.R.~Volkas, \emph{{Phenomenology of a very light
  scalar (100 MeV \ensuremath{<} $m_h$ \ensuremath{<} 10 GeV) mixing with the
  SM Higgs}}, \href{https://doi.org/10.1007/JHEP02(2014)123}{\emph{JHEP}
  {\bfseries 02} (2014) 123} [\href{https://arxiv.org/abs/1310.8042}{{\ttfamily
  1310.8042}}].

\bibitem{SHiP:2018xqw}
{\scshape SHiP} collaboration, \emph{{Sensitivity of the SHiP experiment to
  Heavy Neutral Leptons}},
  \href{https://doi.org/10.1007/JHEP04(2019)077}{\emph{JHEP} {\bfseries 04}
  (2019) 077} [\href{https://arxiv.org/abs/1811.00930}{{\ttfamily
  1811.00930}}].

\bibitem{SHiP:2023}
C.e.a.~Ahdida, \emph{{Post-LS3 Experimental Options in ECN3}}, .

\bibitem{SHiP:ECN3}
{\scshape SHiP} collaboration, \emph{{BDF/SHiP at the ECN3 high-intensity beam
  facility}}, .

\bibitem{Ruf:2115534}
{\scshape SHiPcollaboration} collaboration, \emph{{Heavy Flavour Cascade
  Production in a Beam Dump}}, .

\bibitem{Budnev:1975poe}
V.M.~Budnev, I.F.~Ginzburg, G.V.~Meledin and V.G.~Serbo, \emph{{The Two photon
  particle production mechanism. Physical problems. Applications. Equivalent
  photon approximation}},
  \href{https://doi.org/10.1016/0370-1573(75)90009-5}{\emph{Phys. Rept.}
  {\bfseries 15} (1975) 181}.

\bibitem{Dobrich:2015jyk}
B.~D\"obrich, J.~Jaeckel, F.~Kahlhoefer, A.~Ringwald and K.~Schmidt-Hoberg,
  \emph{{ALPtraum: ALP production in proton beam dump experiments}},
  \href{https://doi.org/10.1007/JHEP02(2016)018}{\emph{JHEP} {\bfseries 02}
  (2016) 018} [\href{https://arxiv.org/abs/1512.03069}{{\ttfamily
  1512.03069}}].

\bibitem{Aloni:2019ruo}
D.~Aloni, C.~Fanelli, Y.~Soreq and M.~Williams, \emph{{Photoproduction of
  Axionlike Particles}},
  \href{https://doi.org/10.1103/PhysRevLett.123.071801}{\emph{Phys. Rev. Lett.}
  {\bfseries 123} (2019) 071801}
  [\href{https://arxiv.org/abs/1903.03586}{{\ttfamily 1903.03586}}].

\bibitem{Harland-Lang:2019zur}
L.~Harland-Lang, J.~Jaeckel and M.~Spannowsky, \emph{{A fresh look at ALP
  searches in fixed target experiments}},
  \href{https://doi.org/10.1016/j.physletb.2019.04.045}{\emph{Phys. Lett. B}
  {\bfseries 793} (2019) 281}
  [\href{https://arxiv.org/abs/1902.04878}{{\ttfamily 1902.04878}}].

\bibitem{Dusaev:2020gxi}
R.R.~Dusaev, D.V.~Kirpichnikov and M.M.~Kirsanov, \emph{{Photoproduction of
  axionlike particles in the NA64 experiment}},
  \href{https://doi.org/10.1103/PhysRevD.102.055018}{\emph{Phys. Rev. D}
  {\bfseries 102} (2020) 055018}
  [\href{https://arxiv.org/abs/2004.04469}{{\ttfamily 2004.04469}}].

\bibitem{Balkin:2023gya}
R.~Balkin, O.~Hen, W.~Li, H.~Liu, T.~Ma, Y.~Soreq et~al., \emph{{Probing
  axion-like particles at the Electron-Ion Collider}},
  \href{https://doi.org/10.1007/JHEP02(2024)123}{\emph{JHEP} {\bfseries 02}
  (2024) 123} [\href{https://arxiv.org/abs/2310.08827}{{\ttfamily
  2310.08827}}].

\bibitem{Wu:2024fsf}
Q.-f.~Wu and X.-J.~Xu, \emph{{A comprehensive calculation of the Primakoff
  process and the solar axion flux}},
  \href{https://arxiv.org/abs/2402.16083}{{\ttfamily 2402.16083}}.

\bibitem{Liu:2016mqv}
Y.-S.~Liu, D.~McKeen and G.A.~Miller, \emph{{Validity of the
  Weizs\"acker-Williams approximation and the analysis of beam dump
  experiments: Production of a new scalar boson}},
  \href{https://doi.org/10.1103/PhysRevD.95.036010}{\emph{Phys. Rev. D}
  {\bfseries 95} (2017) 036010}
  [\href{https://arxiv.org/abs/1609.06781}{{\ttfamily 1609.06781}}].

\bibitem{Liu:2017htz}
Y.-S.~Liu and G.A.~Miller, \emph{{Validity of the Weizs\"acker-Williams
  approximation and the analysis of beam dump experiments: Production of an
  axion, a dark photon, or a new axial-vector boson}},
  \href{https://doi.org/10.1103/PhysRevD.96.016004}{\emph{Phys. Rev. D}
  {\bfseries 96} (2017) 016004}
  [\href{https://arxiv.org/abs/1705.01633}{{\ttfamily 1705.01633}}].

\bibitem{Vysotskii:2018eic}
M.I.~Vysotskii and E.~Zhemchugov, \emph{{Equivalent photons in proton and
  ion\textemdash{}ion collisions at the LHC}},
  \href{https://doi.org/10.3367/UFNe.2018.07.038389}{\emph{Phys. Usp.}
  {\bfseries 62} (2019) 910}
  [\href{https://arxiv.org/abs/1806.07238}{{\ttfamily 1806.07238}}].

\bibitem{Ma:2021lgv}
Z.-L.~Ma, Z.~Lu and L.~Zhang, \emph{{Validity of equivalent photon spectra and
  the photoproduction processes in p-p collisions}},
  \href{https://doi.org/10.1016/j.nuclphysb.2021.115645}{\emph{Nucl. Phys. B}
  {\bfseries 974} (2022) 115645}
  [\href{https://arxiv.org/abs/2112.14399}{{\ttfamily 2112.14399}}].

\end{thebibliography}\endgroup
\end{document}